\documentclass[aps,prb,twocolumn,superscriptaddress,showpacs]{revtex4-1}
\usepackage{longtable}
\usepackage{supertabular}
\usepackage{graphicx}
\usepackage{amsfonts}
\usepackage{mathrsfs}

\begin{document}

\title{Cluster packing geometry for Al-based F-type icosahedral alloys}

\author{Nobuhisa Fujita}
\email[]{nobuhisa@tagen.tohoku.ac.jp}
\affiliation{Institute of Multidisciplinary Research for Advanced Materials, Tohoku University, Sendai 980-8577, Japan}
\author{Hikari Takano}
\affiliation{Institute of Multidisciplinary Research for Advanced Materials, Tohoku University, Sendai 980-8577, Japan}
\author{Akiji Yamamoto}
\affiliation{National Institute for Materials Science, Tsukuba 305-0044, Japan}
\author{An-Pang Tsai}
\affiliation{Institute of Multidisciplinary Research for Advanced Materials, Tohoku University, Sendai 980-8577, Japan}
\affiliation{National Institute for Materials Science, Tsukuba 305-0044, Japan}

\date{\today}

\begin{abstract}
This paper presents a new highly stable periodic approximant to the Al-based F-type icosahedral quasicrystals, $i$-Al-Pd-TM (TM=transition metals). The structure of this intermetallic Al-Pd-Cr-Fe compound is determined {\it ab initio} using single-crystal X-ray diffraction, where the space group is identified to be Pa$\bar{3}$ and the lattice constant 40.5\AA. The structure is well described as a dense packing of clusters of two kinds, which are known in the literature as the pseudo-Mackay type and the Bergman type clusters. The clusters are centered at the vertices of a canonical cell tiling, in which the parity of each vertex determines the kind of the associated cluster. Adjacent clusters can be markedly interpenetrated, while the structure requires no glue atoms to fill in the gaps between the clusters. It is shown that the crystal can be designated as a $2\times2\times2$ superstructure of the ordinary cubic 3/2 rational approximant. The superlattice ordering is shown to be of a different kind from the P-type superlattice ordering previously reported in $i$-Al-Pd-Mn. The present results will greatly improve the understanding of atomic structures of F-type icosahedral quasicrystals and their approximants. 
\end{abstract}


\maketitle                        


\section{Introduction}

The Al based F-type icosahedral quasicrystals, such as {\em i}-Al$_{65}$Cu$_{20}$Fe$_{15}$ and {\em i}-Al$_{70}$Pd$_{20}$TM$_{10}$ (TM=transition metal, e.g., Mn, Re), pose a number of questions as to the detailed atomic arrangements. Researchers have tried to extract as much structural information as possible from materials of this kind through state-of-the-art techniques for structure analysis using X-ray, electron or neutron diffraction. These studies indicated that there existed a few different kinds of clusters as the basic building units that construct the quasicrystals.

A direct measurement of a quasicrystalline sample, however, entails a well known disadvantage in terms of structure determination. Conventional methods developed for analyzing crystal structures are inherently inapplicable to quasicrystals due to the absence of periodicity. However, these methods could still provide invaluable information regarding the constituent clusters as well as their local packing in quasicrystals through analyzing their rational approximants. In particular, the latter approach has so far been taken successfully in elucidating local characteristics of P-type icosahedral quasicrystals, such as {\em i}-Al$_{73}$Mn$_{21}$Si$_6$ \cite{elserhenley85,duneauoguey89} and {\em i}-Cd$_{5.7}$Yb \cite{takakura07}.

Unfortunately, for the case of F-type icosahedral quasicrystals, a stable approximant phase has scarcely been reported. Hence, local characteristics of the structure remained to be uncertain to a large extent. It appears as if Al-based F-type quasicrystals are so stable that no subordinate approximant phase could be obtained via a slight change of the composition. To date, only two possibly related cubic approximants have been previously reported in a rapidly quenched alloy with a nominal composition of Al$_{69}$Pd$_{20}$Mn$_{8}$Si$_{3}$ after annealing at above $1000^\circ$C \cite{ishimasa92}. By way of single crystal X-ray diffraction, these approximants were determined to be cubic $1/1$ and $2/1$ approximants to the F-type icosahedral quasicrystal based on their lattice constants of 12.28\AA~ and 20.21\AA~ \cite{ksugiyama98_1by1,ksugiyama98_2by1}. The structural information of the two refined structures has been utilized as the basis for building a six-dimensional structure model of the relevant F-type icosahedral quasicrystal \cite{yamamoto03}. However, the reliability of the two structures has remained controversial because their constituent clusters were totally different despite their compositional similarity.

In {\em i}-Al$_{70}$Pd$_{20}$TM$_{10}$ and its close associates, Al, Pd and TM are likely to maintain their distinct roles in the structure. While no favorable approximant to the icosahedral quasicrystal has been reported for TM = Mn or Re, an ample room remains for the choice of TM elements. In particular, a fine adjustment of the electron concentration (e.g., e/a ratio) could be achieved by mixing two transition metal elements for TM. In this report, a synthesis of a new stable approximant is accomplished by blending Cr and Fe, the two immediate neighbors to Mn in the periodic table, for TM. An optimal ratio between Cr and Fe is searched experimentally. The stability of the new phase is such that fine single crystals may grow through a simple slow cooling method. Single crystal X-ray diffraction is then performed to analyze the structure. The first part of this paper is devoted to a presentation of the synthesis and the crystal structure analysis.

The present structure analysis offers a source of information that is essential in understanding the local atomic arrangements in the F-type icosahedral quasicrystals. It turns out that the structure is composed of two kinds of clusters, called the pseudo-Mackay type and Bergman type clusters. The clusters are interconnected in such a way that significant interpenetrations are allowed, while no glue atom between them is required. Moreover, the centers of the clusters are given as the nodes (or vertices) of a three-dimensional tiling with four kinds of polyhedra called the canonical cells \cite{clhenley91}; this finding is of fundamental importance as it allows a systematic description of the basic skeleton of this as well as other related compounds. The main body of the work comprises a full account of the crystal structure and related discussions.

The structural description can be generalized in a straightforward manner, leading to the proposal of a number of hypothetical structures which could form as real approximants. Although a full description of the F-type icosahedral quasicrystals awaits future endeavors, our understanding on the local atomic arrangements in the F-type icosahedral quasicrystals and their approximants can now be revised significantly based on firm experimental evidences. This yet opens a way to further attempt to synthesize variants of stable approximants as well as to gain insights on the superlattice ordering phenomenon which was reported previously in the F-type icosahedral quasicrystal {\em i}-Al$_{70}$Pd$_{20}$TM$_{10}$ \cite{ishimasa95}.

This paper is organized as follows. Section \S\ref{sec2} illustrates how a simple compositional search culminates in the discovery of a new approximant phase. The initial samples were polycrystalline, yet powder X-ray diffraction as well as electron diffraction proves the presence of a cubic approximant with a large unit cell. An electron micro-probe analyzer (EPMA) is used to evaluate the fine chemical composition of the approximant. Then an optimized synthesis is performed, resulting in the growth of single crystals exceeding 100 $\mu m$ in diameter. Single crystal X-ray diffraction and the structure analysis are performed in Section \S\ref{sec3}. The crystal structure is described in detail in Section \S\ref{sec4} in a constructive manner: After introducing the geometrical templates of the two kinds of clusters, the packing geometry of the clusters is described in detail. It is shown that the global arrangement of the clusters is determined based on a canonical cell tiling with F-type ordering. In an early part of Section \S\ref{sec5}, discussions are given of the relationship between a superlattice ordering of the present approximant and that in the quasicrystal. A likelihood of anti-phase boundaries as a possible source of disorder in the present material is also discussed. Then the rest of this section examines the possibility of introducing a tiling model for the atomic arrangement. Section \S\ref{sec6} is devoted to concluding remarks.

\section{Synthesis and characterization}\label{sec2}

Our synthesis of the new approximant was performed in two steps. Firstly, a search for a possible approximant phase was made by varying the ratio between Cr and Fe. A few characterization techniques were used to locate the composition of the approximant phase as well as to extract the basic crystallographic information. Secondly, starting from the right composition a new synthesis was performed with a slow cooling. High quality single crystals of the approximant were successfully obtained in this way. These steps are described in the following.

The starting materials were Al (Kojundo Chemical Lab.; purity, 99.9\%), Pd (Tanaka Kikinzoku; purity, 99.95\%), Cr (Furuuchi Chemical; purity, 99.9\%) and Fe (Nilaco; purity, 99.5\%). An alloy ingot with a nominal composition of Al$_{70}$Pd$_{20}$Cr$_x$Fe$_{10-x}$ ($x=3$, 5 or 7) was prepared by the arc-melting method under argon atmosphere. After sufficient homogenization was achieved, the as-solidified ingot was fragmented, put into a Tammann crucible made of Al$_2$O$_3$ and sealed altogether into a quartz tube with pure argon gas of about 0.08 MPa. In an electric furnace, the sample was annealed at $850^\circ$C for 48 hours and cooled swiftly ($\sim$ 1 hour) down to room temperature.

\begin{figure}
\caption{\label{fig:1} Powder X-ray diffraction diagrams for Al$_{70}$Pd$_{20}$Cr$_x$Fe$_{10-x}$ ($x=3$, 5 or 7). For $x=3$ and 5, the peaks marked with filled circles are associated with an approximant, while for $x=7$ those marked with filled squares are associated with a quasicrystal. Note that unassigned peaks associated with  impurity phases become prominent as $x$ is increased}
\includegraphics[width=8cm]{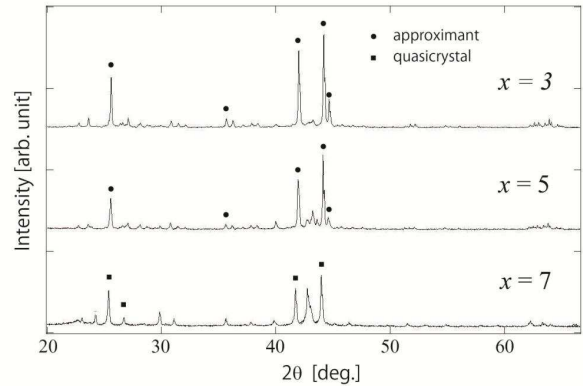}
\end{figure}

In Figure \ref{fig:1}, powder X-ray diffraction patterns taken from the samples $x=3$, 5 and 7 are shown, where the characteristic X-ray of Cu K$\alpha$ ($\lambda=1.543$\AA) is used with Bragg-Brentano diffraction geometry (Mac Science, diffractometer M03XHF$^{22}$). Slightly above the strongest Bragg peak at $2\theta\sim 44^\circ$, a clear sub-peak is observed for the two samples $x=3$ and 5. The latter peak feature is usually associated with the cubic 2/1 approximant to an icosahedral quasicrystal with the Miller indices being {\em hkl}=10 0 0. The corresponding lattice constant is calculated to be c.a. $20.3$\AA. The sample $x=3$ appears to contain the largest amount of the approximant phase, while the suppression of the sub-peak as well as the growth of an extra feature at $2\theta\sim 43^\circ$ clearly indicates the prevalence of impurity phase(s) as $x$ is increased. Hence, it is expected that Al$_{70}$Pd$_{20}$Cr$_3$Fe$_7$ is the closest composition to that of the approximant phase.

\begin{figure}
\caption{\label{fig:2}SEM back-scattered electron images for (a) $x$$=$$3$, (b) 5 and (c) 7. The estimated compositions are: (a) Al$_{69.1}$Pd$_{22.0}$Cr$_{2.1}$Fe$_{6.8}$ (light gray) and Al$_{68.4}$Pd$_{20.9}$Cr$_{1.1}$Fe$_{9.5}$ (dark gray), (b) Al$_{69.7}$Pd$_{22.5}$Cr$_{2.3}$Fe$_{5.4}$ (light gray) and Al$_{72.5}$Pd$_{11.8}$Cr$_{11.6}$Fe$_{4.0}$ (dark gray) and (c) Al$_{69.5}$Pd$_{23.5}$Cr$_{3.6}$Fe$_{3.4}$ (light gray) and Al$_{3}$Pd$_{2}$ (white).}
\includegraphics[width=6cm]{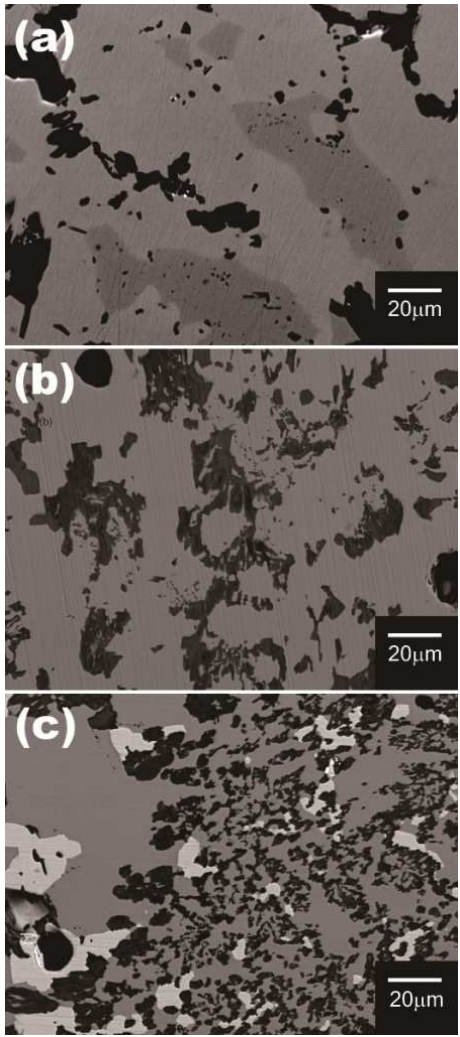}%
\end{figure}

Back-scattered electron images taken from the three samples using a scanning electron microscope (SEM) are shown in Figure \ref{fig:2}. The SEM apparatus used is JXA-8621MX (JEOL), in which an electron-probe micro-analyzer (EPMA) is implemented for chemical composition analysis. A few domains of different gray levels are observed in each sample (voids are shown as black regions), indicating the existence of a few alloy phases with different compositions. It is observed that a light gray region in Figure \ref{fig:2}(a) represents the majority phase for $x=3$ with the estimated composition being Al$_{69.1}$Pd$_{22.0}$Cr$_{2.1}$Fe$_{6.8}$. It is reasonable to assume that this region corresponds to the approximant, which constitute the main body of the sample. For $x=5$, a light gray region in Figure \ref{fig:2}(b) has an estimated composition of Al$_{69.7}$Pd$_{22.5}$Cr$_{2.3}$Fe$_{5.4}$, and one can still associate it with the approximant for the compositional difference is insubstantial. However, all the phases observed for $x=7$ are deviated significantly from the approximant in terms of chemical composition, implying no relevant approximant phase exists in the sample. 

\begin{figure}
\caption{\label{fig:3}TEM diffraction patterns taken from the Al$_{70}$Pd$_{20}$Cr$_3$Fe$_{7}$ sample along (a) the two-fold axis ($\parallel$ $<100>$), (b) the three-fold axis ($\parallel$ $<111>$) and (c) one of the pseudo five-fold axes.}
\includegraphics[width=6cm]{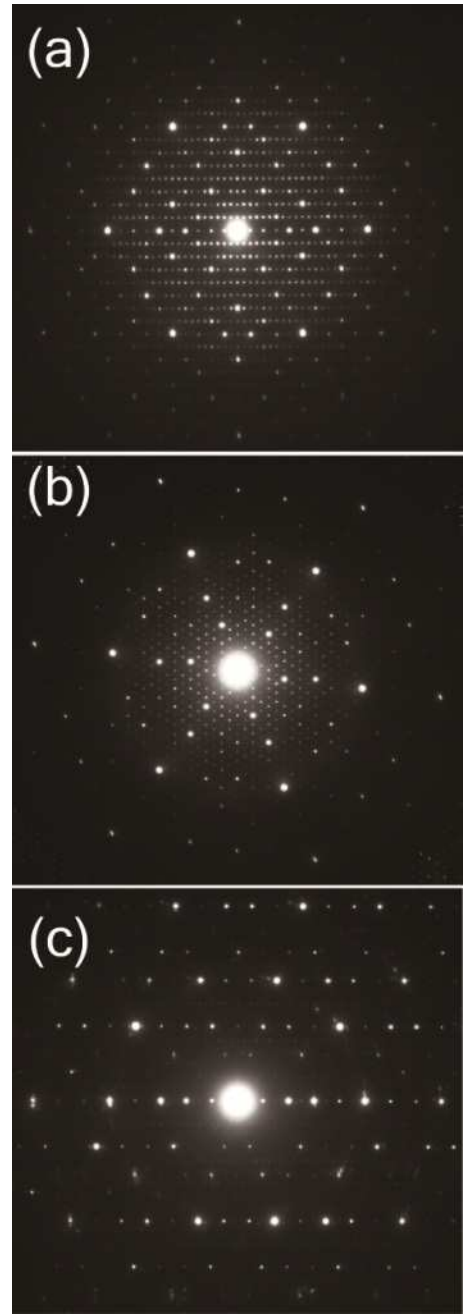}%
\end{figure}

Firmer evidence of the approximant phase was provided by selected-area electron diffraction. Here two transmission electron microscopes (TEM), JEM-2000EX (JEOL) and JEM-2000EXII (JEOL), operating at 200 kV were used. Electron diffraction patterns taken from the Al$_{70}$Pd$_{20}$Cr$_3$Fe$_{7}$ sample (Figure \ref{fig:3}) clearly indicate the existence of 2-fold, 3-fold and pseudo 5-fold axes, where strong Bragg reflections are arranged in a similar way to the case of the quasicrystal \cite{aptsai90}. In Figure \ref{fig:3}(a) and (b), periodic arrays of spots can be indexed as a cubic crystal with a lattice constant of about $40$\AA. Note that, however, this is twice as large as the one estimated from the above powder X-ray diffraction pattern by assuming a conventional 2/1 approximant. Moreover, extinctions are observed at 0{\em kl} with {\em k}=odd and at 00{\em l} with {\em l}=odd, indicating that the crystal has a non-symmorphic space group. Therefore, the pre-existing model of a 2/1 approximant \cite{ksugiyama98_2by1,dmitrichizhi07} clearly fails to describe the present material. This motivates us to carry out a thorough investigation of the crystal geometry based on an ab initio structure determination using single crystal X-ray diffraction.

\begin{figure}
\caption{\label{fig:4} SEM micrograph taken from the single crystalline sample obtained with a nominal composition of Al$_{69.4}$Pd$_{22.3}$Cr$_{2.2}$Fe$_{6.1}$.}
\includegraphics[width=8cm]{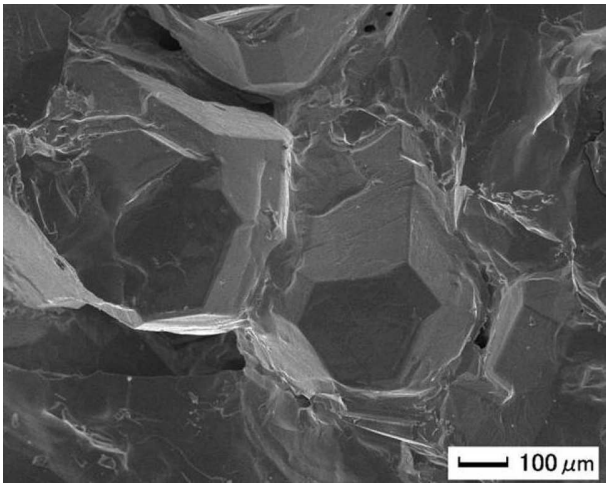}%
\end{figure}

In order to obtain a single crystal to be used for the structure analysis, a new sample was prepared in the following way. A refined composition given by the arithmetic mean, Al$_{69.4}$Pd$_{22.3}$Cr$_{2.2}$Fe$_{6.1}$, of the two compositions evaluated from the samples $x=3$ and 5 was used for the starting composition. The preparation steps were the same as those described above except that at the annealing step within the furnace the sample was fully melted at  $1160^\circ$C for one hour and then cooled down slowly to room temperature with a cooling rate of $10^\circ$C/hour. Figure \ref{fig:4} shows an SEM micrograph taken from the new sample using JCM-5100 (JEOL), where faceted crystalline grains roughly in the shape of a truncated octahedron are wrapped partially with debris (impurities). It confirms that single crystals a few hundred $\mu m$'s in diameter were successfully grown, which also implies that the approximant formed congruently from the melt.

\section{Ab initio structure determination}\label{sec3}

\subsection{Single crystal X-ray diffraction}

A single crystal was taken from the new sample with the estimated dimensions of 0.272{\em mm} $\times$ 0.178{\em mm} $\times$ 0.320{\em mm}. The diffraction experiments were performed using a Bruker SMART APEX diffractometer, mounting CCD area detector, with Mo K$\alpha$ radiation and graphite monochromator (wave length $=0.71073$\AA). Indexing and empirical absorption correction were performed using the Bruker software package (SMART for WNT/2000 5.625 Bruker AXS and SAINT 6.45 SADABS). In collecting the reflection intensities, the crystal was identified as a primitive cubic crystal with a lattice constant of $40.54$\AA. 

Inspecting the original reflection data, it was found that a set of symmetrically equivalent reflections often included one or a few members showing exceptional deviation in the intensity. This is due to the dynamical effect (or multiple scattering) by which an extra intensity is added to the `kinematical' Bragg intensity. Although the extra intensities are weak in contrast to those of the strongest Bragg reflections, they could still be harmful for weak reflections; note that the structural information carried by weak reflections plays a crucial role in determining long-range characteristics of a complex structure such as a quasicrystal and an approximant with a large unit cell.

In principle, a reflection intensity affected by multiple scattering could be effectively screened out from the data based on the intensity distribution among the set of equivalent {\it hkl} entries. In a common structure refinement software (e.g., JANA2006), the procedure can be performed when equivalent reflections are averaged. However, if the number of equivalent {\it hkl} entries is small, the statistics is simply not enough to discern the ill-entry. Hence, we took a precaution to remove those {\it hkl} entries which have less than six symmetrically equivalent associates from the original dataset in order to avoid the possible source of error. The preprocessed dataset was used for the structure analysis as described in the following subsection.

\subsection{Structure analysis}

Our structure analysis relied entirely on the JANA2006 software package \cite{jana2006}. The Laue group and the space group was determined unambiguously to be $m\bar{3}$ ($T_h$) and $Pa\bar{3}$, which turned out to be the only non-symmorphic space group with minor contradictions with the data. Out of the whole 95213 observed reflections satisfying $I>3\sigma(I)$, 302 symmetrically extinct reflections were included. The symmetry averaging was performed while at the same time those data entries showing significant deviation from the averages of their equivalent reflections ($|I-I_{av}|>10\sigma(I_{av})$) were eliminated from the data. The averaged data containing 8454 independent reflections satisfying $I>3\sigma(I)$ were obtained with $R_{int}=7.91\%$.

The initial structure model was generated with the SUPERFLIP program, which is an implementation of the charge flipping algorithm in JANA2006. Then Fourier synthesis and least squares fitting were iterated. After several iterations, the automated peak search turned unsuccessful in locating new atoms. Then we carefully inspected the tentative structure, finding incomplete icosahedral clusters orderly packed in the unit cell. We identified two kinds of clusters, which in the literature are known as the pseudo-Mackay type clusters and the Bergman type clusters. It was observed that adjacent clusters were connected along either a two-fold or a three-fold symmetry axis of the reference icosahedron; a two-fold linkage connects clusters of the same kind and a three-fold linkage connects ones of different kinds. After the central atoms of all the clusters were identified, the missing atoms within each cluster were located by inspecting the charge density plot. Finally the refinement converged with the reliability index of $R=11.96$ (\%) or $R_w=12.13$ (\%). A detailed presentation of the refinement as well as a table of atomic parameters is provided in Appendix \S\ref{appendixB}. The basic crystallographic data are summarized in Table \ref{crystaldata}.

\begin{table}
\caption{Crystallographic data\label{crystaldata}}
\begin{center}
\begin{tabular}{ll}
Formula & Al$_{72.515}$ Pd$_{22.498}$ Cr$_{4.928}$ Fe$_{7.853}$\\
Molar mass & 5045.2  g/mol\\
Temp. of data collection & room temp.\\
Space group & Pa$\bar{3}$ (No. 205)\\
Lattice constant, $a_{lat}$ & 40.5405 \AA\\
Cell volume, $\Omega$ & 66629.6 \AA$^3$\\
$Z$ & 40\\
Calculated density & 5.028  g/cm$^3$\\
Absorption coefficient & 9.294 mm$^{-1}$ \\
Range of 2$\theta$ &  1.74 $\sim$ 53.2 $^\circ$\\
Independent reflections & 21526\\
Obs. reflections ($I>3\sigma(I)$) & 8454\\
$R_{\rm int}$ (obs/all) & 7.91/13.37 \\
Num. of parameters & 783 \\
$R(F)$ & 11.96\\
$R_w(F)$ & 12.13\\
$S$ & 4.59\\
$\Delta\rho_{\rm max}$, $\Delta\rho_{\rm min}$  & $8.01$, $-5.38$ e/\AA$^3$\\
$\Delta/$e.s.d. &  0.0002\\
\end{tabular}
\end{center}
\end{table}

\section{Crystal structure}\label{sec4}

The refined structure contains 4728 atomic sites per unit cell, wherein 204 sites are symmetrically unique. Importantly, the structure can be pictured as one formed through dense packing of clusters, which are allowed to overlap with each other across their peripheries. There is no need for glue atoms to fill in the gaps between the clusters, meaning each of the atomic sites belongs to at least one cluster. Two kinds of cluster called the pseudo-Mackay type and the Bergman type clusters are identified; these are henceforth referred to as M- and B-clusters, respectively.

\subsection{Cluster templates}

\begin{figure}
\caption{\label{fig:5}The idealized templates for the two kinds of cluster. (a) M-cluster consists of a central site (M$_0$), a dodecahedral shell (M$_3$) and a composite shell with an icosahedral subshell (M$_5$) and an icosidodecahedral subshell (M$_2$). (b) B-cluster consists of a central site (B$_0$), an icosahedral shell (B$_5$) and a dodecahedral shell (B$_3$).}
\includegraphics[width=8cm]{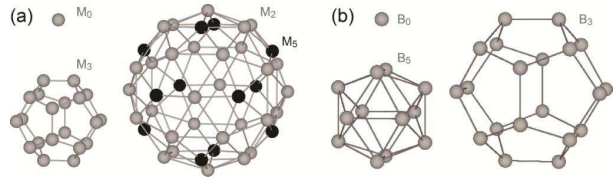}%
\end{figure}

The geometrical templates for M- and B-clusters are shown in Figure \ref{fig:5}; both have an atomic site at the center and two shells having the full icosahedral symmetry, $\bar{5}\bar{3}2/m$ ($I_{\rm h}$). The central sites are symbolized as M$_0$ and B$_0$ for M-cluster and B-cluster, respectively. The inner shell of M-cluster consists of twenty sites forming the vertices of a regular dodecahedron. This shell is symbolized as M$_3$ because the relevant sites are on the three-fold rotation axes of the icosahedral point group; the same principle also applies when symbolizing the remaining shells. The outer shell of M-cluster is a composite of two subshells, twelve sites forming a regular icosahedron (M$_5$) and thirty sites forming a regular icosidodecahedron (M$_2$). On the other hand, the inner shell of B-cluster consists of twelve sites forming a regular icosahedron (B$_5$), while the outer shell twenty sites forming a regular dodecahedron (B$_3$).

Let us now define the six icosahedral basis vectors ${\bf a}_j$ ($j=1$, 2, ... and 6) as
\begin{eqnarray}
&&\left(\begin{array}{cccccc}
{\bf a}_1 & {\bf a}_2 & {\bf a}_3 & {\bf a}_4 & {\bf a}_5 & {\bf a}_6
\end{array}\right)\nonumber\\
&&:= \ell\left(\begin{array}{cccccc}
 \tau &    0 &    1 &     1 & -\tau &     0 \\
    1 & \tau &    0 &     0 &     1 & -\tau \\
    0 &    1 & \tau & -\tau &     0 &     1 
\end{array}\right),\label{physbasis}
\end{eqnarray}
where $\tau=(1+\sqrt{5})/2$ is the golden mean and $\ell$ represents an appropriate scale. The norm of the basis vectors is hereafter denoted $a (:=|{\bf a}_j|)$. Note that the numbering, $j=1$, 2, ... and 6, of the six basis vectors is taken such that they are arranged concentrically around a three-fold rotation axis \footnote{Another numbering convention that is faithful to a five-fold rotation axis \cite{velser85} is more common in the literature.}.
Taking the central site of a cluster template as the origin, the position vector {\bf x} of every site within the cluster template is written as ${\bf x}=\sum_j c_j {\bf a}_j=:[c_1c_2c_3c_4c_5c_6]$, in which the indices $c_j$ ($j=1$, 2, ... and 6) are all integers or half-integers. To be more specific, each of the shells, M$_x$ and B$_x$ ($x=0, 2, 3$ or $5$), is obtained as an orbit of its representative member with respect to the point group $\bar{5}\bar{3}2/m$ ($I_{\rm h}$); the representative indices and the radius are listed in Table \ref{table:shellslinkages}.

\begin{table}
\caption{\label{table:shellslinkages} The indices given for each shell represent the vector from the center to a representative site of the shell. Representative vectors for b- and c-linkages as well as the lattice translation vectors are also given for reference. The norm of each vector is given in the right column. Note that $b=2(\tau^3/\sqrt{5})^{1/2}a$ and $b=\sqrt{3}/2 c$. Negative indices are represented by integers with top bars.}
\begin{center}
\begin{tabular}{cll}
shell & indices & norm \\ \\
M$_0$ & $[0 0 0 0 0 0]$ & $0$\\
M$_2$ & $[1 1 0 0 0 0]$ & $b/\tau$\\
M$_3$ & $[1 1 1 1 1 1]/2$ & $c/\tau^2$\\
M$_5$ & $[1 1 1 1 \bar{1} \bar{1}]/2$ & $\tau a$\\
B$_0$ & $[0 0 0 0 0 0]$ & $0$\\
B$_3$ & $[\bar{1} \bar{1} \bar{1} 1 1 1]/2$ & $c/\tau$\\
B$_5$ & $[1 0 0 0 0 0]$ & $a~ (=1.902\ell)$\\ \\
b-linkage & $[1 1 0 0 1 \bar{1}]$ &  $b~ (=5.236\ell)$\\
c-linkage & $[1 1 1 0 0 0]$ & $c~ (=4.535\ell)$\\
${\bf R}_1$ & $[6 0 4 4 \bar{6} 0]$ & $2b\tau^2$\\
${\bf R}_2$ & $[4 6 0 0 4 \bar{6}]$ & $2b\tau^2$\\
${\bf R}_3$ & $[0 4 6 \bar{6} 0 4]$ & $2b\tau^2$\\
\end{tabular}
\end{center}
\end{table}

\subsection{Packing geometry}\label{subsec:packinggeometry}

In the refined structure (Appendix \S\ref{appendixB}), the unit cell accommodates a total of 264 clusters, which divide into 128 M- and 136 B-clusters. In Table \ref{table:clustercenters}, the independent positions of the cluster centers are summarized. The centers of the M-clusters (M$_0$) are occupied mainly by Fe or Cr, although better fit is attained by occupying some of them partially with Pd or Al. The centers of the B-clusters (B$_0$) are purely occupied by Pd.

\begin{table}
\caption{\label{table:clustercenters}Centers of the clusters within the unit cell extracted from the refined parameters (Appendix \S\ref{appendixB}).}
\begin{center}
\begin{tabular}{ccccc}
site symbol & atom & Wyckoff & indices & node type \\ \\
B$^{(1)}_0$&Pd1 & $8c$ & $[1  1  1  0  0  0]$ &   $(67)_{333}$ \\
B$^{(2)}_0$&Pd6 & $8c$ & $[4  4  4 \bar{1} \bar{1} \bar{1}]$ &  $(67)_{333}$ \\
B$^{(3)}_0$&Pd11 & $24d$ & $[3  3  2  0  0 \bar{1}]$ &  $(66)_{4322}$ \\
B$^{(4)}_0$&Pd24 & $24d$ & $[4  4  3  0  0 \bar{2}]$ &  $(67)_{333}$ \\
B$^{(5)}_0$&Pd37 & $24d$ & $[3  1  2  1 \bar{2}  0]$ &   $(76)_{433}$ \\
B$^{(6)}_0$&Pd50 & $24d$ & $[4  2  3  1 \bar{2} \bar{1}]$ &   $(76)_{433}$ \\
B$^{(7)}_0$&Pd63 & $24d$ & $[5  3  3  1 \bar{1} \bar{2}]$ &   $(67)_{333}$ \\
M$^{(1)}_0$&Cr/Al76 & $4a$ & $[0  0  0  0  0  0]$ &  $(68)_0$ \\
M$^{(2)}_0$&Cr/Al79 & $4b$ & $[3  0  2  2 \bar{3}  0]$ &  $(68)_0$ \\
M$^{(3)}_0$&Fe/Pd82 & $24d$ & $[4  3  2  1  0 \bar{2}]$ &  $(57)_{3322}$ \\
M$^{(4)}_0$&Fe/Pd90 & $24d$ & $[4  1  2  2 \bar{2} \bar{1}]$ &   $(57)_{332}$ \\
M$^{(5)}_0$&Cr/Al98 & $24d$ & $[2  2  1  0  0 \bar{1}]$ &   $(66)^{\prime}_{432}$ \\
M$^{(6)}_0$&Fe/Pd105 & $24d$ & $[4  5  1  0  2 \bar{4}]$ &  $(67)_{333}$ \\
M$^{(7)}_0$&Fe113 & $24d$ & $[4  3  3  0 \bar{1} \bar{1}]$ &  $(67)_{333}$ \\
\end{tabular}
\end{center}
\end{table}

Adjacent clusters are mutually connected either along a two-fold or a three-fold axis with a distance of about 7.7\AA~ or 6.7\AA, respectively; these are the shortest two distances between cluster centers. An analogous feature, with the distances being appropriately scaled, has been extensively discussed in the case of P-type icosahedral quasicrystals and their approximants \cite{clhenley86,clhenley91,mihalkovic96}, in which only a single kind of cluster would come into play. The two- and three-fold linkages are called b- and c-linkages, respectively, whereas a packing of clusters with this property is called a bc-packing. The b- and c-linkages are indexed with integers and their representatives are given in Table \ref{table:shellslinkages}; the respective norms are denoted $b$ and $c$. The skeleton of the crystal structure can thus be described as a network of nodes, which correspond to the cluster centers, connected through b- and c-linkages.

\begin{figure}
\caption{\label{fig:6}The three different combinations for an adjacent pair of clusters: M-M (left), B-B (middle) and M-B (right). Two clusters of the same kind are connected via b-linkage, while those of the two different kinds are connected via c-linkage. The sites in the constituent cluster templates are depicted as spheres. The drawing plane is perpendicular to a 2-fold symmetry axis.}
\includegraphics[width=8cm]{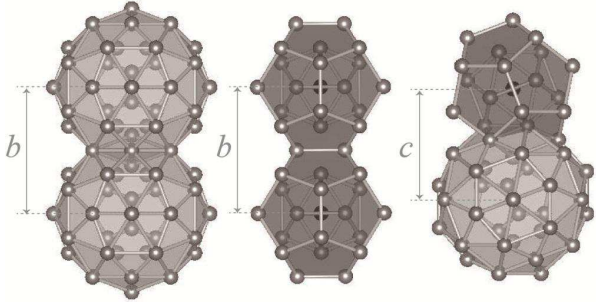}%
\end{figure}

F-type ordering (or F-centering) in icosahedral quasicrystals as well as their approximants occurs as the even and odd parities of the nodes in the relevant bc-packing are differentiated. Here, the parity of each node is defined as that of the sum of the relevant indices. Note in Table \ref{table:shellslinkages} that each b-linkage connects a pair of nodes having the same parity, while that each c-linkage connects a pair of nodes having different parities. Therefore, if the two kinds of cluster, M and B, are the entities that differentiate the two subsets of the nodes of a bc-packing, it follows that every b-linkage connects the same kind of cluster (M-M or B-B), while that every c-linkage connects different kinds of cluster (M-B). The three combinations for an adjacent pair of clusters are depicted in Figure \ref{fig:6}.

The coordinates of the cluster centers given in the refined atomic parameters (Appendix \S\ref{appendixB}) can be used to enumerate which of the b- and c-linkages connect every site to its adjacent neighbors. The task of obtaining the indices for every cluster center is straightforward. First we set the indices of the atomic site Cr/Al76 lying at the origin (0,0,0), which is the center of an M-cluster, to $[0 0 0 0 0 0]$. Then we recursively trace linkages to obtain the indices of adjacent cluster centers until all the cluster centers within the unit cell as well as the lattice translation vectors are identified. The resulting lattice translation vectors ${\bf R}_j$ ($j=1$, 2 and 3) are presented in Table \ref{table:shellslinkages}. Since the lattice constant, $|{\bf R}_j|=2b\tau^2$ should equal the experimental value of $a_{lat}=40.54$\AA, the basic parameter $b$ is immediately evaluated as 7.74\AA.

Now we are equipped with an appropriate scale for examining the inter-atomic distances. Let us first note that the ideal edge length $b/\tau^3$ of the inner dodecahedral shell of M-cluster (M$_3$; Figure \ref{fig:5}(a)) is evaluated to be $1.83$\AA, and that it is unrealistically short for an interatomic distance. A natural consequence of this is that the relevant shells cannot be occupied by more than 8 atoms at the same time. Next, note that the outer shell of M-cluster (including the two subshells) has two kinds of edges. The edges connecting the adjacent pairs in the M$_2$ subshell are parallel to the two-fold axes and have a length of $b^\prime:=b/\tau^2=2.958$\AA, while those connecting the M$_2$ subshell with the M$_5$ subshell are parallel to the three-fold axes and have a length of $c^\prime:=c/\tau^2=2.562$\AA. The latter two edge lengths are reasonable for interatomic distances. Similarly, the edges connecting the adjacent pairs in the inner (B$_5$) or the outer (B$_3$) shell of B-cluster are parallel to the two-fold axes and have a length of $b^\prime$. Interestingly, the two kinds of linkages, namely $b^\prime$ and $c^\prime$, form the majority of the interatomic linkages in the idealized construction. This applies not only to the closest interatomic linkages within each shell of a cluster but also to those connecting between the inner and the outer shells. It follows that the closest distances are $b^\prime$ for the pairs, M$_3$-M$_3$, M$_2$-M$_2$, M$_3$-M$_5$, B$_5$-B$_5$ and B$_3$-B$_3$, while $c^\prime$ for the pairs M$_2$-M$_5$, M$_3$-M$_2$ and B$_5$-B$_3$. Here we have included the pair M$_3$-M$_3$ because the closest interatomic distance that is allowed within the partially occupied inner shell (M$_3$) of an M-cluster is $b^\prime$.

Two M-clusters connected through b-linkage interpenetrate into each other (Figure \ref{fig:6}, left) with the overlap between the outer polyhedra being a flat hexagonal bipyramid. The two tips of the bipyramid belong to the M$_2$ shells and are very close to each other (1.828\AA), so that they cannot be occupied at the same time. In the refinement, such positions have been fitted as splitting positions for a single Al atom. The intersection between the two M-clusters consists of the six sites forming the base hexagon of the hexagonal bipyramid; two of them belong to the M$_5$ subshells of the two M-clusters, while four of them to the M$_2$ subshells.
On the other hand, two B-clusters connected through b-linkage share an edge between their outer shells (B$_3$, Figure \ref{fig:6}, middle).
In both cases, the intersection between the two clusters does not involve the inner shells.

The situation is somewhat more intricate in the case of an M-B pair connected through c-linkage. Observe in Figure \ref{fig:6} (right) that the outer shell (B$_3$) of the B-cluster penetrates to the inner shell (M$_3$) of the M-cluster, where a site is shared by these shells.
Similarly, the outer subshell (M$_2$) of the M-cluster penetrates to the inner shell (B$_5$) of the B-cluster, where a triangular face is shared by these shells.
One also finds that these clusters intersects with each other in their outer (sub)shells M$_5$ and B$_3$, too, where there are three common sites between the two.

\begin{figure}
\caption{\label{fig:7}The packing of clusters. The square drawn with white dashed lines indicate a face of the cubic unit cell, whereas the horizontal and vertical edges correspond to the primitive lattice vectors, {\bf R}$_1$ and {\bf R}$_2$, respectively.}
\includegraphics[width=8cm]{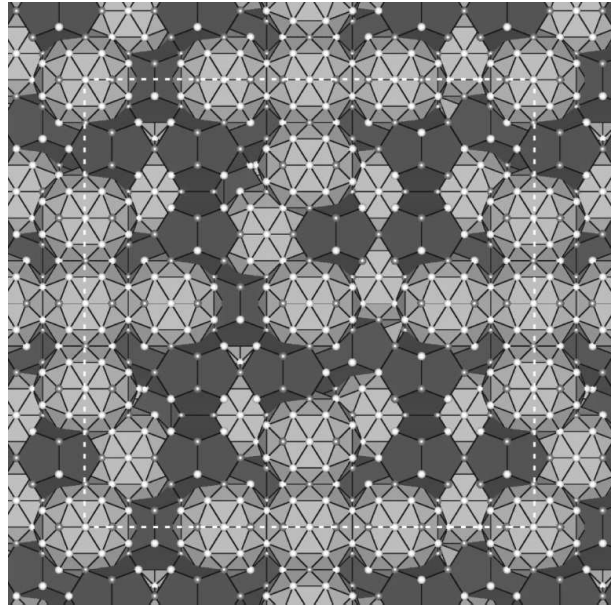}
\end{figure}

The idealized atomic positions can be obtained by replicating the relevant cluster templates at the positions given in Table \ref{table:clustercenters}. The arrangement of clusters in a unit cell is illustrated in Figure \ref{fig:7}. As we have just seen, an atomic position can be shared by two or more cluster shells that intersect with each other. Therefore, each individual site can be characterized by the set of the cluster shells to which it belongs. Take, for instance, a site in the inner shell (M$_3$) of an M-cluster belonging also to the outer shell (B$_3$) of an adjacent B-cluster. We simply assign to this site a class, $\langle$M$_3$,B$_3\rangle$. More generally, if an atomic site belongs to $n$ cluster shells, $X^1, X^2, ..., X^n$, then the class associated with this site would be $\langle X^1,X^2, ...,X^n\rangle$. Bear in mind that only the combination of the cluster shells matters here, so that the order of the shell symbols in the angle brackets are irrelevant. It turns out that the idealized structure of the present crystal contains in total 16 different classes of atomic sites, which are listed in Table \ref{table:sitesymbols}.

\begin{table}
\caption{\label{table:sitesymbols} The 16 different classes of idealized atomic sites. Their number frequencies per unit cell, presented in the second column, sum up to a total of 4680. The average composition (in percentage) of each class is given in the last column, where `Vc' stands for vacancy.}
\begin{center}
\begin{tabular}{lll}
site class & number & average composition\\ \\
$\langle$M$_0\rangle$ & 128 ($=F$) & Al$_{6}$Pd$_{4}$(Cr,Fe)$_{90}$\\
$\langle$M$_2\rangle$ & 720 ($=G$) & Al$_{50}$Vc$_{50}$\\
$\langle$M$_3\rangle$ & 1680 ($=H$) & Vc$_{100}$\\
$\langle$B$_0\rangle$ & 136 ($=I$) & Pd$_{100}$\\
$\langle$B$_5\rangle$ & 24 ($=J$) & Al$_{100}$\\
$\langle$M$_2$,M$_2\rangle$ & 72 ($=K$)  & Al$_{100}$\\
$\langle$M$_2$,B$_5\rangle$ & 576 ($=L$) & Al$_{100}$\\
$\langle$M$_3$,B$_3\rangle$ & 880 ($=M$) & Al$_{81}$Pd$_{1}$(Cr,Fe)$_{17}$Vc$_{1}$\\
$\langle$M$_5$,B$_3\rangle$ & 24 ($=N$) & Pd$_{100}$\\
$\langle$B$_3$,B$_3\rangle$ & 72 ($=O$) & Pd$_{100}$\\
$\langle$M$_2$,M$_2$,B$_5\rangle$ & 1032 ($=P$) & Al$_{100}$\\
$\langle$M$_2$,M$_2$,M$_2\rangle$ & 112 ($=Q$) & Al$_{100}$\\
$\langle$M$_5$,M$_5$,M$_5\rangle$ & 24 ($=R$) & Al$_{1}$(Cr,Fe)$_{99}$\\
$\langle$M$_5$,B$_3$,B$_3\rangle$ & 408 ($=S$) & Pd$_{98}$(Cr,Fe)$_{2}$\\
$\langle$M$_5$,M$_5$,B$_3$,B$_3\rangle$ & 384 ($=T$) & Pd$_{65}$(Cr,Fe)$_{35}$\\
$\langle$M$_5$,M$_5$,M$_5$,B$_3\rangle$ & 88 ($=U$) & Pd$_{3}$(Cr,Fe)$_{97}$\\
\end{tabular}
\end{center}
\end{table}

It is obvious from the above argument that the central site of each cluster cannot be shared by any other cluster. Hence, the symbols for the cluster centers are $\langle$M$_0\rangle$ and $\langle$B$_0\rangle$. In addition, the crystal structure includes three more classes of unshared sites, which are $\langle$M$_2\rangle$,$\langle$M$_3\rangle$ and $\langle$B$_5\rangle$. Importantly, the former two bear distinct roles in the physical construction of the structure: (i) $\langle$M$_2\rangle$, the splitting positions given at the tip of the overlap hexagonal bipyramid associated with each interpenetrating pair of M-clusters and (ii) $\langle$M$_3\rangle$, the vacant sites in the M$_3$ shells of the M-clusters. Remember that the inner shell of M-cluster (M$_3$) cannot accommodate more than 8 atoms due to the constraint imposed by the short nearest-neighbor distance. And it is clearly demonstrated by our structure analysis that the only occupied M$_3$ sites are those represented by the symbol $\langle$M$_3$,B$_3\rangle$. The last unshared class $\langle$B$_5\rangle$ does not seem to differ from the other types of B$_5$ sites, as all of them are fully occupied by Al. 

Correlations are further found between the local compositions at individual atomic sites and the site classes. Take, for instance, the B$_3$ shell, which involves six different site classes (Table \ref{table:sitesymbols}). Whereas the sites symbolized as $\langle$M$_3$,B$_3\rangle$ are primarily occupied by Al, the sites in the remaining five classes, $\langle$M$_5$,B$_3\rangle$, $\langle$B$_3$,B$_3\rangle$, $\langle$M$_5$,B$_3$,B$_3\rangle$, $\langle$M$_5$,M$_5$,B$_3$,B$_3\rangle$ and $\langle$M$_5$,M$_5$,M$_5$,B$_3\rangle$, are occupied by markedly heavier elements in different degrees depending on their site classes. Similar observations suggest that the physical nature of an individual atomic site is profoundly affected by the manner how clusters intersect there. Therefore, the two kinds of clusters not only provide a handy means of describing the complex structure but also serve as the true physical units which play a significant role in the formation of the structure. This underlying basic idea may also apply to a more general class of Al-based alloys which exhibit F-type icosahedral ordering.

The packing of the cluster templates as described above generates in total 4680 idealized atomic positions (excluding the vacant $\langle$M$_3\rangle$ sites), which agree remarkably well with the refined atomic positions, although in the refinement we included 48 additional sites which are needed to explain minor irregularities in the real material. Importantly, the gaps between the clusters are small enough and do not accommodate any additional atoms (called glue atoms). In Subsection \S\S\ref{subsec:atomicpacking}, this feature is described from a somewhat different viewpoint.

\subsection{Canonical cell tiling with F-type ordering}\label{subsec:canonicalcells}

Recall that the atomic arrangement and the chemical compositions within each individual cluster are subject to constraints posed by the existence of adjacent clusters (Subsection \S\S\ref{subsec:packinggeometry}). These geometrical constraints may hardly allow other combinations of clusters than the three cases shown in Figure \ref{fig:6}, thus enforcing a perfect F-type ordering. The F-type icosahedral ordering observed in more general Al-based alloys could also have a similar geometrical origin, where an uneven distribution of atomic species on different sublattices is understood to be a secondary consequence.

\begin{figure}
\caption{\label{fig:8}The canonical cells and their geometrical components. (a) The two kinds of linkages, b and c, represented as thick and double bars, respectively. (b) The three kinds of faces, X, Y and Z, whose sides are distinguishable in terms of the icosahedral symmetry; where `$+$' sign is on one side, `$-$' sign is on the other side. The reverse sides of the faces are shown by rotating them by 180 degrees around the vertical dashed line. (c) The four canonical cells, A, B, C and D, and a trigonal anti-prism unit, B$_2$, which can be divided into two B-cells in three ways. The unique corners are indicated by the corner symbols, while the numbers on the edges correspond to the following dihedral angles: (1) $\pi/2$, (2) $\pi/3$, (3) $\pi-\epsilon$, (4) $\pi-\eta$, (5) $\eta$, (6) $\pi-2\eta$, (7) $(\pi-\epsilon)/2$, (8) $\epsilon$, (9) $\eta$, (10) $\pi/2$ and (11) $\pi/3$.}
\includegraphics[width=8cm]{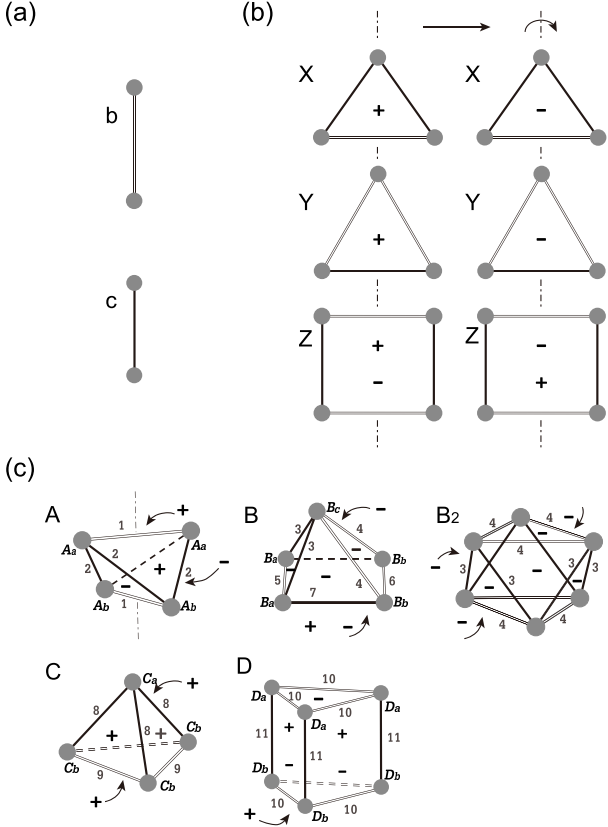}%
\end{figure}

\begin{figure}
\caption{\label{fig:9} (color) A top view of a cleaved surface of the present CCT parallel to the (001) plane; the foremost cells are chosen somewhat arbitrarily. The spheres representing the even (resp. odd) nodes are colored white (resp. black). The white dashed lines indicate a square face of the cubic unit cell, while the horizontal and vertical edges correspond to the primitive lattice vectors, {\bf R}$_1$ and {\bf R}$_2$, respectively. At the lower left corner of the square lies an even node corresponding to the origin. The four kinds of canonical cells are colored yellow (A-cell), red (B-cell), blue (C-cell) and green (D-cell), respectively.}
\includegraphics[width=8cm]{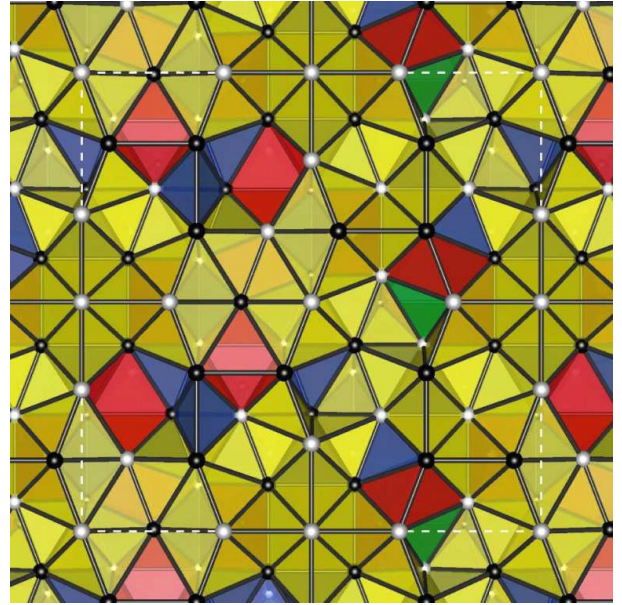}%
\end{figure}

The skeletal structure of the present approximant is described as the bc-packing, in which the cluster centers are represented as the nodes (Table \ref{table:clustercenters}). The nodes can be connected to each other through edges of the two kinds (b- and c-linkages; Figure \ref{fig:8}(a)), whereas the edges form three kinds of polygons: an isosceles triangle formed by a b-linkage and two c-linkages (X-face; point symmetry, $m$), an equilateral triangle formed by three b-linkages (Y-face; $3m$) and a rectangle formed by two b-linkages and two c-linkages (Z-face; $2/m$); Figure \ref{fig:8}(b). These polygons are further found to be the faces of four kinds of polyhedra called {\it the canonical cells} \cite{clhenley91}: a tetrahedron with four X-faces (A-cell; $2m$), a pyramid with three X-faces, one Y-face and one Z-face (B-cell; $m$), a tetrahedron with three X-faces and one Y-face (C-cell; $3m$) and a trigonal prism with two Y-faces and three Z-faces (D-cell; $3m$); Figure \ref{fig:8}(c). The present bc-packing thus proves to be represented as a periodic tiling of space called a canonical cell tiling (CCT); that is, the whole space is divided into pieces congruent to the canonical cells (Figure \ref{fig:9}).

So far, suitable atomic decorations of periodic CCT's have been commonly used to describe a variety of approximant phases to P-type icosahedral quasicrystals \cite{clhenley91,mihalkovic96}. On the other hand, the present compound demonstrates for the first time that the CCT construction extends naturally to approximant phases to F-type icosahedral quasicrystals. Here, the two parities of the vertices need to be clearly distinguished to account for the F-type ordering; this is done by representing even (resp. odd) vertices with white (resp. black) spheres in Figure \ref{fig:9}. The geometrical composition for the present CCT in particular is summarized in Table \ref{table:numberofcells}. There are 11 congruence classes of objects, some of which are further divided into two subclasses (I and II) owing to the parities of the nodes, leading to the 18 object classes in total. Remember that X-face as well as A-, B-, C- or D-cell has more than one unique corners (see the corner symbols in Table \ref{table:numberofcells}), for which the parities need to be specified explicitly. It follows that there are two ways in coloring the vertices of each kind of cell, leading to two different configurations of clusters and thus to two different atomic arrangements within the cell (in the sense that the atomic species are disregarded).

\begin{table}
\caption{\label{table:numberofcells} The number frequencies of the geometrical objects contained in a single unit cell of the present CCT. The 11 congruent types can be further divided into subclasses according to the parities of the nodes as specified in the parentheses using the symbols, $+$ (for even) and $-$ (for odd), leading to 18 basic object types. Note that for the faces (X - Z) as well as the cells (A - D), the parity symbols are arranged in the same order as the corner symbols referenced in the second column. The object type B$_2$ can be divided into two B-cells in three different ways, and we do not take it as a basic object type.}
\begin{center}
\begin{tabular}{lll}
object & \multicolumn{2}{l}{type}\\
           &  (I) & (II) \\ \\
node & 128 ($+$) & 136 ($-$) \\
edge b & 360 ($+\:+$) & 432 ($-\:-$)\\
edge c & \multicolumn{2}{l}{* 880 ($+\:-$)}\\
face X (X$_a$X$_a$X$_b$) & 1032 ($++-$) & 1176 ($--+$)\\
face Y (Y$_a$Y$_a$Y$_a$) & 112 ($+++$) & 176 ($---$)\\
face Z (Z$_a$Z$_a$Z$_a$Z$_a$) & \multicolumn{2}{l}{* 176 ($++-\:-$)}\\
cell A (A$_a$A$_a$A$_b$A$_b$) & 384 ($++-\:-$) & 336 ($--+\:+$)\\
cell B (B$_a$B$_a$B$_b$B$_b$B$_c$) & 104 ($--+++$) & 152 ($++---$)\\
cell B$_2$ (B$_c$B$_c$B$_c$B$_c$B$_c$B$_c$) & \multicolumn{2}{l}{* 80 ($+++--\:-$)}\\
cell C  (C$_a$C$_b$C$_b$C$_b$) & 88 ($-++\:+$) & 168 ($+--\:-$)\\
cell D (D$_a$D$_a$D$_a$D$_b$D$_b$D$_b$) & 8 ($+++--\:-$) & 24 ($---++\:+$)\\
\end{tabular}
\end{center}
\end{table}

The present argument extends naturally to arbitrary CCT's, whereby a number of hypothetical approximants with F-type ordering can be constructed. From this broader perspective, it is worthwhile making a general consideration on the statistics of the geometrical objects as well as that of the atomic sites. Still, readers who might want to avoid handling the mathematical derivations may safely skip the rest of this subsection and go to Section \S\ref{sec5}.

In the following argument, the number density of an arbitrary object type $O$ among the 18 basic object types is denoted $n(O)$. In particular, the number densities of the eight types of cells are of fundamental importance, so that we introduce the following parameters: $\alpha_{\rm I}=n({\rm A_{I}})$, $\alpha_{\rm II}=n({\rm A_{II}})$, $\beta_{\rm I}=n({\rm B_{I}})$, $\beta_{\rm II}=n({\rm B_{II}})$, $\gamma_{\rm I}=n({\rm C_{I}})$, $\gamma_{\rm II}=n({\rm C_{II}})$, $\delta_{\rm I}=n({\rm D_{I}})$ and $\delta_{II}=n({\rm D_{II}})$. The following sums are also used whenever it is convenient to do so:
\begin{eqnarray}
\alpha&:=&\alpha_{\rm I}+\alpha_{\rm II},\quad \beta:=\beta_{\rm I}+\beta_{\rm II},\nonumber\\
\gamma&:=&\gamma_{\rm I}+\gamma_{\rm II},\quad \delta:=\delta_{\rm I}+\delta_{\rm II}.
\end{eqnarray}
By definition, the sum rule for the cell volumes is written as
\begin{eqnarray}
v_A \alpha + v_B \beta + v_C \gamma + v_D \delta &=& 1,\label{eq:volumesum}
\end{eqnarray}
where $v_A=b^3/12$, $v_B=\sqrt{5}b^3/12$, $v_C=\sqrt{5}b^3/24$ and $v_D=3b^3/8$ are the volumes of A-, B-, C- and D-cells, respectively. In addition to this, these statistical parameters hold a set of universal equations due to geometrical constraints imposed by the shapes of the cells.

Let us first consider the matching constraints across the faces. Take, for instance, a cell that has an X$_{\rm I}$ face with its $+$ side facing outward, then another cell with an X$_{\rm I}$ face with its $-$ side facing outward can only match the former cell. Since the same requirement is fulfilled across any of the X$_{\rm I}$ faces in the structure, the $+$ and $-$ sides of the faces of this type must appear equal number of times. By counting the two distinct sides of X$_{\rm I}$ faces for each type of cells, Eq.(\ref{eqXI}) can be readily verified. Furthermore, similar considerations when applied to the X$_{\rm II}$, Y$_{\rm I}$ and Y$_{\rm II}$ faces lead to Eqs.(\ref{eqXII}), (\ref{eqYI}) and (\ref{eqYII}), respectively.
\begin{eqnarray}
2\alpha_{\rm I}+3\gamma_{\rm I} &=& 2\alpha_{\rm II}+2\beta_{\rm I}+\beta_{\rm II},\label{eqXI}\\
2\alpha_{\rm II}+3\gamma_{\rm II} &=& 2\alpha_{\rm I}+2\beta_{\rm II}+\beta_{\rm I},\label{eqXII}\\
\beta_{\rm I}+\delta_{\rm I} &=& \gamma_{\rm I}+\delta_{\rm II},\label{eqYI}\\
\beta_{\rm II}+\delta_{\rm II} &=& \gamma_{\rm II}+\delta_{\rm I}.\label{eqYII}
\end{eqnarray}

Now we show that the matching constraints around the edges will lead to two more equations. Note that the dihedral angles contained in a B- or C-cell (see Figure \ref{fig:8}(c)) involve two irrational constant, $\epsilon$ and $\eta$, defined by \cite{clhenley91}
\begin{eqnarray}
\epsilon&\equiv&\arccos{(\frac{1}{4})}\approx 4\pi(0.1049),\\
\eta&\equiv&\arccos{(\frac{1}{\sqrt{6}})}\approx 4\pi(0.0915).
\end{eqnarray}
Here each cell contributes a fixed amount of dihedral angles associated with a fixed edge type. Hence, the contributions from all the cells sum up to the total number of the relevant edges times $2\pi$; that is,
\begin{eqnarray}
2\pi n({\rm b_{I}})&=&(\pi/2)\alpha+(3\pi-4\eta)\beta_{\rm I}+\eta\beta_{\rm II}\nonumber\\
&&+3\eta\gamma_{\rm I}+(3\pi/2)\delta,\label{2piNbI}\\
2\pi n({\rm b_{II}})&=&(\pi/2)\alpha+\eta\beta_{\rm I}+(3\pi-4\eta)\beta_{\rm II}\nonumber\\
&&+3\eta\gamma_{\rm II}+(3\pi/2)\delta,\label{2piNbII}\\
2\pi n({\rm c})&=&(4\pi/3)\alpha+3(\pi-\epsilon)\beta+3\epsilon\gamma+\pi\delta.\label{2piNc}
\end{eqnarray}
Observe that in Eqs.(\ref{2piNbI})-(\ref{2piNc}) terms containing $\epsilon$ and $\eta$ do not cancel out spontaneously. Hence, in order to avoid unphysical consequences that is the number of edges of each type would be irrational, we need to constrain the coefficients for $\epsilon$ and $\eta$ to be zero. Therefore, the following equations need to be satisfied;
\begin{eqnarray}
-4\beta_{\rm I}+\beta_{\rm II}+3\gamma_{\rm I}&=&0,\quad \beta_{\rm I}-4\beta_{\rm II}+3\gamma_{\rm II} = 0,\nonumber\\
-3\beta+3\gamma&=&0.\label{eqepseta}
\end{eqnarray}

One can readily check that Eqs.(\ref{eqXI})-(\ref{eqYII}) and (\ref{eqepseta}) can be reduced to the following four universal equations,
\begin{eqnarray}
\beta&=&\gamma,\quad \Delta\alpha=-\Delta\beta,\nonumber\\
3\Delta\gamma&=&5\Delta\beta,\quad 3\Delta\delta=\Delta\beta,\label{eq:shapeconstraint}
\end{eqnarray}
where
\begin{eqnarray}
\Delta\alpha&:=&\alpha_I-\alpha_{II},\quad \Delta\beta:=\beta_I-\beta_{II},\nonumber\\
\Delta\gamma&:=&\gamma_I-\gamma_{II},\quad \Delta\delta:=\delta_I-\delta_{II}.
\end{eqnarray}

It follows from Eqs.(\ref{eq:volumesum}) and (\ref{eq:shapeconstraint}) that there are only three independent degrees of freedom to determine the statistical properties of any CCT with F-type ordering. Following ref.\onlinecite{clhenley91}, an independent parameter $\mu$ is defined as the volume fraction occupied by A- and D-cells:
\begin{eqnarray}
\mu &:=& v_A \alpha + v_D \delta, 
\end{eqnarray}
while another $\zeta$ is defined so that $\mu\zeta$ gives the volume fraction occupied by D-cells:
\begin{eqnarray}
\zeta :=& v_D \delta/\mu.
\end{eqnarray}
It is readily shown that these two parameters determine the number densities of the canonical cells through \cite{clhenley91}
\begin{eqnarray}
\alpha &=& 12\mu (1-\zeta)/b^3,\\
\beta &\equiv& \gamma = (8/\sqrt{5}) (1-\mu) /b^3,\\
\delta &=&  (8/3) \mu \zeta/b^3.
\end{eqnarray}
The last independent parameter $\nu$, which is necessary if F-type ordering is considered, can be defined as
\begin{eqnarray}
\nu &:=& \Delta\beta/\beta.
\end{eqnarray}

As soon as the statistics is fixed for the cells, it is a straightforward task to enumerate the objects of lower dimensionalities (faces, edges and nodes) by taking the contributions from all the cells. The edges have already been enumerated via Eqs.(\ref{2piNbI})-(\ref{2piNc}), while for enumerating the faces the double counting of each face must be taken care of. The nodes are enumerated by taking the sum of the solid angles associated with the relevant corners of the cells \cite{clhenley91} and by dividing the results by $4\pi$. The resulting formulae are
\begin{eqnarray}
n({\rm node_I})&=&\alpha/12+\beta/4+\Delta\beta/12+\delta/4,\\
n({\rm node_{II}})&=&\alpha/12+\beta/4-\Delta\beta/12+\delta/4,\\
n({\rm b_I})&=&\alpha/4+3(\beta+\Delta\beta)/4+3\delta/4,\\
n({\rm b_{II}})&=&\alpha/4+3(\beta-\Delta\beta)/4+3\delta/4,\\
n({\rm c})&=&2\alpha/3+3\beta/2+\delta/2,\\
n({\rm X_I})&=&\alpha+3(\beta+\Delta\beta)/2,\\
n({\rm X_{II}})&=&\alpha+3(\beta-\Delta\beta)/2,\\
n({\rm Y_I})&=&\beta/2+2\Delta\beta/3+\delta/2,\\
n({\rm Y_{II}})&=&\beta/2-2\Delta\beta/3+\delta/2,\\
n({\rm Z})&=&(\beta+3\delta)/2.
\end{eqnarray}
The numbers of objects listed in Table \ref{table:numberofcells} are the frequencies within a single unit cell. These values can be readily compared with the above formulae by re-interpreting the variables as the frequencies per unit cell, that is, $\alpha\Omega=720$, $\beta\Omega=256$, $\Delta\beta\Omega=-48$ and $\delta\Omega=32$. 

Each of the above 18 object types offers a possible local atomic configuration through replicating the relevant cluster templates. Therefore, the above statistics uniquely determines the number densities $f=F/\Omega$, $g=G/\Omega$, ..., $u=U/\Omega$ of the 16 types of atomic sites; see Table \ref{table:sitesymbols}. One can readily prove the following general formulae (see Appendix \S\ref{appendix:sitestat}):
\begin{eqnarray}
f &=& 1/12\alpha+1/4\beta+1/12\Delta\beta+1/4\delta,\label{eq_sitestat_i}\\
g &=& 1/2\alpha+3/2\beta+3/2\Delta\beta+3/2\delta,\\
h &=& \alpha+7/2\beta+5/3\Delta\beta+9/2\delta,\\
i &=& 1/12\alpha+1/4\beta-1/12\Delta\beta+1/4\delta,\\
j &=& 1/2\Delta\beta+3/2\delta,\\
k &=& -1/2\Delta\beta+3/2\delta,\\
l &=& 3/2\beta-3\Delta\beta+3/2\delta,\\
m &=& 2/3\alpha+3/2\beta+1/2\delta,\\
n &=& 1/2 \Delta\beta+3/2\delta,\\
o &=& -1/2\Delta\beta+3/2\delta,\\
p &=& \alpha+3/2\beta+3/2\Delta\beta,\\
q &=& 1/2\beta+2/3\Delta\beta+1/2\delta,\\
r &=& -1/6\Delta\beta+1/2\delta,\\
s &=& 3/2\beta-1/2\Delta\beta,\\
t &=& 1/2\alpha-1/2\Delta\beta,\\
u &=& 1/2\beta+5/6\Delta\beta.\label{eq_sitestat_f}
\end{eqnarray}
Remembering that each individual site exhibits a chemical composition that is substantially correlated with the site class, the derived statistics could be useful when estimating the chemical composition of a hypothetical crystal that is derived from a CCT. This may facilitate an experimental search for unknown approximants in closely related alloy systems.

\section{Discussion}\label{sec5}

The new model presented in Section \S\ref{sec4} allows a concise description of the extremely complex material, offering the possibility that the stabilization mechanisms as well as the physical properties will be clarified on a structure basis. In particular, the model will lead us to revised views on bondings, defects and dynamics in related materials. At present, our model cannot be extended naively to icosahedral quasicrystals because there is currently no proof as to whether a CCT \cite{clhenley91} can be made quasiperiodic nor having the icosahedral symmetry \footnote{Computational techniques for generating CCT's with large periods have been developed by several authors \cite{mihalkovicmrafko93,mejnewman95}. These techniques search possible arrangements of canonical cells under preset periodic boundary conditions, where the amount of computations will expand exponentially as the periods are increased. There is no way that the preset periods can be extrapolated to infinity; hence, it still remains a challenge to verify the existence of quasiperiodic CCT's.}. One can nevertheless gain an important insight into the quasicrystalline state by evaluating how close (or how far) the given crystal lies to (or from) an idealized quasicrystal. The following subsections are devoted to discussions on some of these topics.

\subsection{Superlattice ordering}\label{subsec:phason}

In the standard six-dimensional formalism for icosahedral quasicrystals, the three-dimensional physical space $\mathbb{E}$ is thought of as a subspace of a six-dimensional hyper-space $\tilde{\mathbb{E}}$, whereas the orthogonal complement to the physical space is called the perpendicular space $\mathbb{E}_{\perp}$. The hyper-space is the direct sum of the two subspaces; i.e., $\tilde{\mathbb{E}}=\mathbb{E}\oplus\mathbb{E}_{\perp}$. Accordingly the basis vectors ${\bf a}_j$ ($\in \mathbb{E}$) defined in Eq.(\ref{physbasis}) are lifted into six dimensions via $\tilde{\bf a}_j=({\bf a}_j, c{\bf a}^\perp_j)$, where $c (\in \mathbb{R})$ is an arbitrary non-zero factor and ${\bf a}^\perp_j$ ($\in \mathbb{E}_\perp$) are defined by
\begin{eqnarray}
&&\left(\begin{array}{cccccc}
{\bf a}^\perp_1 & {\bf a}^\perp_2 & {\bf a}^\perp_3 & {\bf a}^\perp_4 & {\bf a}^\perp_5 & {\bf a}^\perp_6
\end{array}\right)\nonumber\\
 &&:= \ell\left(\begin{array}{cccccc}
1 & 0 & -\tau & -\tau & -1 & 0 \\
-\tau & 1 & 0 & 0 & -\tau & -1 \\
0 & -\tau & 1 & -1 & 0 & -\tau
\end{array}\right).
\end{eqnarray}
If $c$ is taken to be unity, the six-dimensional basis vectors $\tilde{\bf a}_j$ ($\in\tilde\mathbb{E}$) form an orthogonal basis set that generates a six-dimensional hyper-cubic lattice, $\tilde{\mathbb{L}}_{\rm P}$, where the subscript P stands for the primitive lattice (P-type). The norm of the six-dimensional basis vectors are hereafter denoted $\tilde{a} (:=|\tilde{\bf a}_j|=\sqrt{2}a)$, which is evaluated to be 3.977 ~\AA~ from the present approximant.

Provided that we have a structure in which the arrangement of M- and B-clusters is subject to the local rules presented in Subsection \S\S\ref{subsec:packinggeometry}, each cluster is centered at a position that is represented as
\begin{eqnarray}
{\bf x}&=&n_1 {\bf a}_1+n_2 {\bf a}_2+n_3 {\bf a}_3+n_4 {\bf a}_4+n_5 {\bf a}_5+n_6 {\bf a}_6\nonumber\\
&=&[n_1 n_2 n_3 n_4 n_5 n_6],\label{eq.xnj}
\end{eqnarray}
where $n_j$ ($j=1$, 2, ... and 6) are the indices, i.e., the integer coefficients that are determined uniquely if the origin is taken at one of the cluster centers. The kind of the cluster (resp. M or B) depends on the parity of {\bf x} (resp. even or odd), where the parity is defined as that of the sum $\sum_j=n_j$.

The three-dimensional vector {\bf x} can be lifted to a six-dimensional counterpart $\tilde{\bf x}\in\tilde{\mathbb{L}}_{\rm P}$ simply by replacing ${\bf a}_j$ in Eq.(\ref{eq.xnj}) with $\tilde{\bf a}_j$. Then, ${\bf x}$ is the image of $\tilde{\bf x}$ through the orthogonal projection of $\tilde{\mathbb{E}}$ onto $\mathbb{E}$. The image ${\bf x}^\perp$ of $\tilde{\bf x}$ in $\mathbb{E}_\perp$ is definable in a similar way. It is readily understood that the three-dimensional vectors ${\bf x}$ and ${\bf x}^\perp$ maintain a one-to-one correspondence (bijection) with each other, where ${\bf x}^\perp$ is called the conjugate image of ${\bf x}$. The parities of ${\bf x}$, ${\bf x}^\perp$ and $\tilde{\bf x}$ are equal. The even subset (consisting of the vertices of even parity) in $\tilde{\mathbb{L}}_{\rm P}$ forms a sublattice with an index of 2 and it gives an F-type (i.e., F-centered) hyper-cubic lattice $\tilde{\mathbb{L}}_{\rm F}$, having a lattice constant of $\tilde{a}_{\rm F}:=2\tilde{a}$.

The perfect icosahedral symmetry of an ideal quasicrystal requires that the conjugate images of all the cluster centers be bounded within a finite domain $\mathbb{W}$ ($\subset\mathbb{E}_\perp)$ called a window, which is also called an atomic surface or an acceptance domain. If this is the case, the lifted coordinates $\tilde{\bf x}$ are distributed along the three-dimensional cut space $\mathbb{E}+{\bf w}_0$ in $\tilde{\mathbb{E}}$, where ${\bf w}_0 (\in \mathbb{E}_\perp)$ is the center of mass of the window. This can be put also as ${\bf x}^\perp\sim {\bf w}_0$, which is a loose but useful expression in a context where the deviation from the equality is unimportant.

If the cut space is taken to be inclined linearly against $\mathbb{E}$, the coordinates of the cluster centers would behave as ${\bf x}^\perp\sim {\bf A}\cdot{\bf x}+{\bf w}_0$, in which {\bf A} is a $3\times3$ matrix called the linear phason tensor. An approximant in particular has a cut space that is parallel to one of the lattice planes of $\tilde{\mathbb{L}}_{\rm P}$, such that any lattice translation vector {\bf R} for the approximant satisfies ${\bf R}^\perp={\bf A}\cdot{\bf R}$, where ${\bf R}^\perp$ is the conjugate image of ${\bf R}$. Therefore, the task of evaluating {\bf A} for the present approximant can resort to the primitive lattice translation vectors given in Table \ref{table:shellslinkages}, reading
\begin{eqnarray}
{\bf A} &=& \left(\begin{array}{ccc}
{\bf R}^\perp_1 & {\bf R}^\perp_2 & {\bf R}^\perp_3
\end{array}\right)
\left(\begin{array}{ccc}
{\bf R}_1 & {\bf R}_2 & {\bf R}_3
\end{array}\right)^{-1}\nonumber\\
&=&\frac{29 - 13\sqrt{5}}{2}~{\bf I} = -\tau^{-7} ~{\bf I},\label{LPtensorA}
\end{eqnarray}
where {\bf I} is the $3\times3$ identity matrix. The smallness of the linear phason strain implies that the crystal is rather close to an ideal quasicrystal.

\begin{figure}
\caption{\label{fig:10}The packing of clusters for a hypothetical cubic 3/2 approximant. The square drawn with white dashed lines indicates a face of the cubic unit cell, whereas the horizontal and vertical edges correspond to the primitive lattice vectors, ${\bf R}^\prime_1={\bf R}_1/2$ and ${\bf R}^\prime_2={\bf R}_2/2$, respectively.}
\includegraphics[width=8cm]{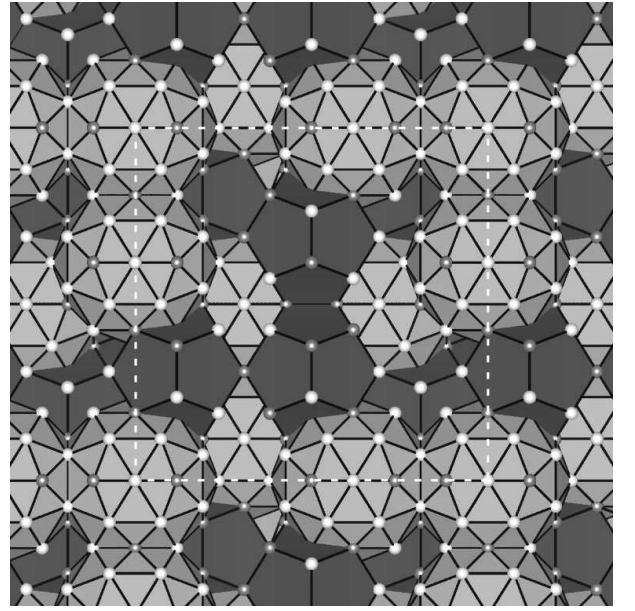}
\end{figure}

The present approximant can indeed be identified as a $2\times2\times2$ superstructure of the cubic 3/2 approximant in the series of rational approximantions of icosahedral quasicrystalinity. The ordinary cubic 3/2 approximant is constructed from a CCT called the 3/2 packing \cite{clhenley91}, which has a lattice constant of $b\tau^2$; then, a hypothetical crystal structure illustrated in Figure \ref{fig:10} is obtained using the decoration rules of Section \S\ref{sec4}. Observe a marked resemblance of this unit cell to each 1/8 block (half cube) cut out from the cubic unit cell of the superstructure (Figure \ref{fig:7}). The primitive lattice translation vectors of the hypothetical 3/2 approximant are ${\bf R}^\prime_j = {\bf R}_j/2$ with $j=1$, 2 and 3. Note that the linear phason tensor is unchanged from that given in Eq.(\ref{LPtensorA}) because the corresponding indices (even numbers) for ${\bf R}_j$ are simply halved for ${\bf R}^\prime_j$. The space group $Pa\bar{3}$ of the 3/2 packing \cite{clhenley91} is reduced to $P2_{1}3$ because the atomic decoration distinguishes the even and odd parities of the nodes. Accordingly, the hypothetical 3/2 approximant should have the enantiomorphic (or chiral) point group $23$. When the superlattice ordering is induced, the space group $Pa\bar{3}$ including the center of symmetry is restored.

It turns out that not only rearrangements of clusters but also additions of extra clusters close to the boundaries between half cubes are necessary in realizing the superstructure. The latter can be checked when the packing densities of the two structures are compared. Observe that a single unit cell of the 3/2 approximant with a volume of $b^3\tau^6$ $(=:\Omega_{3/2})$ contains 16 M- and 16 B-clusters, while that one of the present superstructure with a volume of $\Omega=8 \Omega_{3/2}$ contains 128 M- and 136 B-clusters. Eight additional B-clusters (i.e., $8=136-16\times8$) are included in the unit cell, causing an increase in the number density of clusters from $1.78/b^3$ to $1.84/b^3$. The present superstructure gives a relatively dense packing as compared to most of the simplest CCT's considered by Henley \cite{clhenley91}; the only two with higher densities are the cubic 1/1 and 2/1 packings with the number densities of clusters being $2/b^3$ and $1.89/b^3$, respectively.

\begin{figure}
\caption{\label{fig:11}The modified conjugate images of the cluster centers for the ordinary 3/2 packing (left) and for the $2\times2\times2$ superstructure (right). The black (resp. white) spheres represent the images of the centers of the B-clusters (resp. the M-clusters). The weight (or occupancy) of each modified conjuage image is represented by the volume of the relevant sphere. The truncated octahedron  $\mathscr{O}$ shown in each panel indicates the outer boundary of the fundamental domain in $\mathbb{E}_\perp$; see text for the definition.}
\includegraphics[width=8cm]{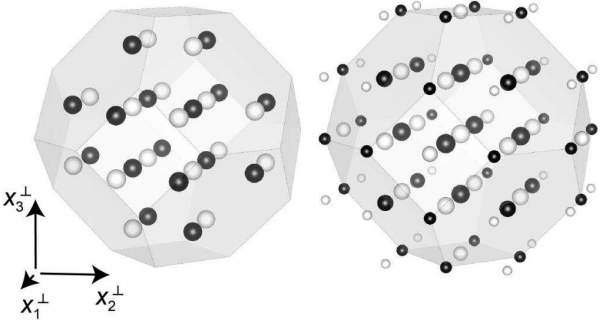}
\end{figure}

A clear contrast between the basic skeleton of the present approximant (superstructure) and that of the 3/2 approximant can be demonstrated with their {\it modified} conjugate images. Generally speaking, the conjugate images of the cluster centers would be unbounded if there exists a non-zero linear phason strain, i.e., ${\bf A} \ne{\bf O}$ ({\bf O}, the $3\times3$ zero matrix), because {\bf x} in ${\bf x}^\perp\sim {\bf A}\cdot{\bf x}+{\bf w}_0$ runs across the physical space $\mathbb{E}$. Hence, one needs to consider the modified conjugate images
\begin{eqnarray}
{\bf x}_{\rm mod}^\perp&:=&{\bf x}^\perp-{\bf A}\cdot{\bf x},\label{modconjimage}
\end{eqnarray}
which behave as ${\bf x}_{\rm mod}^\perp\sim {\bf w}_0$. In Figure \ref{fig:11}, the modified conjugate images of the nodes of the two skeletons are depicted as the white and black spheres, which correspond to the centers of M- and B-clusters, respectively. The truncated octahedron $\mathscr{O}$ depicted in each panel (centered at the center of mass of the modified conjugate images of the nodes) represents to the modified window associated with the 3/2 packing. All the vertices and the hexagonal face centers of  $\mathscr{O}$ are obtained as the modified versions (via Eq.(\ref{modconjimage})) of the 32 vertices of a rhombic triacontahedron $\mathscr{C}_{12}$ ($\subset\mathbb{E}_\perp$), which is the window for the 12-fold vertices in the three-dimensional Penrose tiling; for the definition of $\mathscr{C}_{12}$, refer to the original article \cite{clhenley86} and Figure \ref{fig:9}(a) therein. We find in Figure \ref{fig:11}(left) that the modified conjugate images of the nodes in the 3/2 packing are well confined within $\mathscr{O}$, where every point receives one node per unit cell. On the other hand, those for the $2\times2\times2$ superstructure have wider spread than $\mathscr{O}$ (Figure \ref{fig:11}, right). Importantly, the images lying close to the center of $\mathscr{O}$ receive full number of nodes (8 per unit cell), whereas those in the peripheral region receive less ($<$8). This indicates clearly that the windows associated with the relevant six-dimensional lattice $\tilde\mathbb{L}_{\rm P}$ should be non-uniform.

Relating the present superstructure in $\mathbb{E}$ (3D) to one in $\tilde{\mathbb{E}}$ (6D) is of significant importance in arguing further possibilities that superlattice ordering be observed in other approximants as well as quasicrystals. Indeed, it has been reported by  Ishimasa \cite{ishimasa95} that a superlattice ordering was exhibited by an icosahedral Al-Pd-Mn alloy after being annealed at $700^\circ{\rm C}$. Performing a careful analysis of diffraction patterns, Ishimasa concluded that the Bravais class for the superlattice was P-type, while the corresponding six-dimensional lattice constant was $a_{\rm P}=20.881$\AA$=\tau a_{\rm F}$, where $\tau$ is the golden mean and $a_{\rm F}=12.901$~\AA\, is {\it the conventional six-dimensional lattice constant} for the F-type icosahedral phase Al-Pd-Mn. One can check that the equation $a_{\rm F}=\tau\tilde{a}_{\rm F}$ holds, where $\tilde{a}_{\rm F}$ is the lattice constant of $\tilde{\mathbb{L}}_{\rm F}$; hence, the conventional lattice constant is $\tau$ times that determined in the present study. The discrepancy is however non-essential since there exists an arbitrariness in choosing the basis vectors ${\bf a}_j$ in ${\mathbb{E}}$ up to the multiplication of any integral power of $\tau$ if only given the diffraction module, because of the $\tau$ scaling invariance of the F-type icosahedral module and the corresponding I-type Fourier module \cite{janssen86,rokhsarmermin87,levitovrhyner88}. It is only because the relevant scale for the cluster packing has now been fixed (Subsection \S\S\ref{subsec:packinggeometry}) that we prefer to use a more proper lattice constant $\tilde{a}_{\rm F}$ instead of the conventional one $a_{\rm F}$. The lattice constant for the P-type superlattice ordering in the quasicrystal is therefore written as $a_{\rm P}=\tau^2 \tilde{a}_{\rm F}$. Remember that the use of the updated lattice constant would affect the order of rational approximation; e.g., a conventional cubic 2/1 approximant should now be called a cubic 3/2 approximant.

The simplest P-type sublattice (or superlattice) of $\tilde{\mathbb{L}}_{\rm F}$ is obtained just by taking the subset whose members have only even indices, i.e., $\tilde{\mathbb{L}}^{(0)}_{\rm P}:=2\tilde{\mathbb{L}}_{\rm P}$=$\{[n_1 n_2 n_3 n_4 n_5 n_6];$ $n_j=0$ (mod $2$) for $j=1$, 2, ... and $6\}$. Then it is a sublattice of $\tilde{\mathbb{L}}_{\rm F}$ with an index of $32$ and its six-dimensional lattice constant is given by $\tilde{a}^{(0)}_{\rm P}=\tilde{a}_{\rm F}$. Obviously, $\tilde{\mathbb{L}}^{(0)}_{\rm P}$ fails to represent the superlattice ordering in the quasicrystal because $\tilde{a}^{(0)}_{\rm P}\ne a_{\rm P} (=\tau^2 \tilde{a}_{\rm F})$. In order to obtain the proper sublattice of $\tilde{\mathbb{L}}_{\rm F}$ to describe the superlattice ordering of the quasicrystal, we need first to re-index $\tilde{\mathbb{L}}_{\rm F}$ prior to taking the subset with only even indices. The re-indexing is done with respect to the re-scaled basis vectors $\tilde{\bf a}^{(2)}_j=(\tau^2{\bf a}_j, c(-1/\tau)^2{\bf a}_j^\perp)$, which can be obtained from the original basis vectors $\tilde{\bf a}_j$ through
\begin{eqnarray}
&&(\tilde{\bf a}^{(2)}_1\, \tilde{\bf a}^{(2)}_2\, \tilde{\bf a}^{(2)}_3\, \tilde{\bf a}^{(2)}_4\, \tilde{\bf a}^{(2)}_5\, \tilde{\bf a}^{(2)}_6)\nonumber\\
&&=(\tilde{\bf a}_1\, \tilde{\bf a}_2\, \tilde{\bf a}_3\, \tilde{\bf a}_4\, \tilde{\bf a}_5\, \tilde{\bf a}_6)\,{\bf M}^2,
\end{eqnarray}
where six components for $\tilde{\bf a}_j$ and $\tilde{\bf a}^{(2)}_j$ are all aligned column-wise and {\bf M} represents the uni-modular $\tau$-scale transformation matrix \cite{niizeki89},
\begin{eqnarray}
{\bf M} &=& \frac{1}{2}\left(\begin{array}{cccccc}
1 & 1 & 1 & 1 & -1 & -1\\
1 & 1 & 1 & -1 & 1 & -1\\
1 & 1 & 1 & -1 & -1 & 1\\
1 & -1 & -1 & 1 & -1 & -1\\
-1 & 1 & -1 & -1 & 1 & -1\\
-1 & -1 & 1 & -1 & -1 & 1
\end{array}\right).
\end{eqnarray}
Note that the re-scaled basis vectors $\tilde{\bf a}^{(2)}_j$ are not orthogonal to each other if $c=1$. Nonetheless, since the scale factor $c$ can be arbitrarily chosen without affecting the projected images in $\mathbb{E}$, one can choose $c=\tau^4$ so that the re-scaled basis vectors are orthogonal (i.e., $\tilde{\bf a}^{(2)}_j=\tau^2\tilde{\bf a}_j$) and the sublattice $\tilde{\mathbb{L}}^{(2)}_{\rm P}$ with even indices would regain the hyper-cubic symmetry with a six-dimensional lattice constant of $\tau^2\tilde{a}_{\rm F}$. This proves that Ishimasa's superlattice ordering is associated with the Bravais lattice $\tilde{\mathbb{L}}^{(2)}_{\rm P}$. There can be another kind of P-type sublattice $\tilde{\mathbb{L}}^{(1)}_{\rm P}$ defined after re-indexing with respect to the re-scaled basis vectors $\tilde{\bf a}^{(1)}_j=(\tau{\bf a}_j, c(-1/\tau){\bf a}_j^\perp)$, which are obtained via
\begin{eqnarray}
&&(\tilde{\bf a}^{(1)}_1\, \tilde{\bf a}^{(1)}_2\, \tilde{\bf a}^{(1)}_3\, \tilde{\bf a}^{(1)}_4\, \tilde{\bf a}^{(1)}_5\, \tilde{\bf a}^{(1)}_6) \nonumber\\
&&= (\tilde{\bf a}_1\, \tilde{\bf a}_2\, \tilde{\bf a}_3\, \tilde{\bf a}_4\, \tilde{\bf a}_5\, \tilde{\bf a}_6)\,{\bf M}.
\end{eqnarray}
When the perpendicular space $\mathbb{E}_\perp$ is appropriately scaled (i.e., $c=-\tau^2$), the six-dimensional lattice constant for $\tilde{\mathbb{L}}^{(1)}_{\rm P}$ would be $\tau\tilde{a}_{\rm F}$.

The three sublattices, $\tilde{\mathbb{L}}^{(0)}_{\rm P}$, $\tilde{\mathbb{L}}^{(1)}_{\rm P}$ and $\tilde{\mathbb{L}}^{(2)}_{\rm P}$, of $\tilde{\mathbb{L}}_{\rm F}$ are distinct from each other in terms of translational symmetry. One needs to clarify whether the superlattice ordering in the present approximant can be associated with the same sublattice $\tilde{\mathbb{L}}^{(2)}_{\rm P}$ as in the case of the quasicrystal. In order to simplify the argument, let us focus our attention on the M-cluster whose center lies at the body center ${\bf B}=[555\bar{1}\bar{1}\bar{1}]$ of the conventional unit cell. Obviously, {\bf B} is inequivalent to ${\bf V}=[000000]$, i.e., the vertex of the unit cell, through translation because the crystal is primitive cubic. The latter condition would be satisfied if the relevant superlattice ordering in six dimensions were described by $\tilde{\mathbb{L}}^{(0)}_{\rm P}$ because the indices of {\bf B} are all odd integers and ${\bf B} \notin \tilde{\mathbb{L}}^{(0)}_{\rm P}$$(\ni{\bf V})$. If the superlattice $\tilde{\mathbb{L}}^{(1)}_{\rm P}$ is assumed, the re-scaled indices ${\bf M}^{-1}[555\bar{1}\bar{1}\bar{1}]$ $=$ $[333\bar{1}\bar{1}\bar{1}]$ are again all odd integers, so that ${\bf B} \notin \tilde{\mathbb{L}}^{(1)}_{\rm P}$$(\ni{\bf V})$ is again guaranteed. Yet further re-indexing of {\bf B} leads to ${\bf M}^{-2}[555\bar{1}\bar{1}\bar{1}]$ $=$ $[222000]$, i.e., all even integers, and thence {\bf B} and {\bf V} would become equivalent through translation if the superlattice ordering were described by $\tilde{\mathbb{L}}^{(2)}_{\rm P}$. Therefore, if the superlattice ordering in the approximant had originated from the same physical mechanism as that in the quasicrystal, it would have had body-centered cubic translation symmetry. Perhaps, the superlattice ordering of the approximant can be understood as a kind of lock-in phenomenon facilitated by the discreteness of the atomic surfaces, while for the quasicrystal more subtle competitions between the boundaries of atomic surfaces need to be considered \cite{cockayne94}.

\subsection{Anti-phase boundaries}\label{subsec:antiphaseboundaries}

The atomic decorations of the eight half cubes that comprise the unit cell of the present approximant are mutually congruent in the sense that one of them can be obtained from another through a 2-fold rotation, a mirror reflection or the inversion. This is a strict consequence of the space group $Pa\bar{3}$; in particular, two of the half cubes that share a square face are always connected through a glide operation of the space group $Pa\bar{3}$. Although it is obvious that a half cube does not strictly fall on top of another through a half-way translation, one observes a partial overlap in the atomic positions between the two. It turns out that, from the electron diffraction pattern (Figure \ref{fig:3}) as well as the single crystal X-ray diffraction data, the superlattice reflections (whose Miller indices include at least one odd number) are significantly weaker than the main reflections (whose Miller indices are all even). This indeed justifies the use of the term ``superlattice ordering'' in describing the present structure though topological rearrangements are involved in doubling the translational symmetry, as opposed to a more common type of superlattice orderings associated with density or displacive modulations.

In general, the structure may not be perfectly ideal in realistic materials. For the present alloy sample, in particular, the superlattice ordering can possibly be degraded to some extent due to inequilibrium processes that may have taken place while the material was synthesized. As inspired by the detailed description of the crystal structure (Section \S\ref{sec4}), we expect that the configuration of clusters within each half cube is rather strongly constrained by the local packing rules, whereas the configuration of half cubes is constrained relatively loosely. In this respect, we find that an anti-phase boundary could easily be formed by introducing only a small number of defects in the arrangement of clusters along one of the glide planes \footnote{An anti-phase boundary is a kind of stacking fault at which the lattice planes are shifted by half period. Structural details of anti-phase boundaries in the present material will be presented elsewhere.}. Hence, the material could have exhibited a non-negligible degradation of the superlattice ordering if it were obtained through a non-optimal cooling process. Fortunately, our sample preparation procedure (Section \S\ref{sec2}) yielded a high-quality approximant crystal, in which only minute structural defects are included; hence, no special care was necessary while carrying out the structure refinement.

If the sample preparation were cruder, the superlattice ordering could be violated to a significant extent. Then, the satellite Bragg reflections associated with the superlattice ordering could be seriously damped, possibly making them undetectable. Such a circumstance may have taken place when Sugiyama and coworkers carried out a structure analysis of a closely related Al$_{70}$Pd$_{23}$Mn$_6$Si compound \cite{ksugiyama98_2by1}. These authors reported that their crystal was a cubic 2/1 approximant with a lattice constant of 20.21\AA\, and that the space group was $Pm\bar{3}$. Note that the lattice constant is half times as small as that of the present approximant. Although the reported crystal structure involving a giant icosahedral cluster (diameter, $> 20$\AA) has never indicated a reasonable connection to any other crystals in closely related alloys, the knowledge was transported to the modeling of the icosahedral quasicrystals \cite{yamamoto03}. From our revised view, the halved lattice constant is likely to have resulted from failing to detect the superlattice reflections that suffered from a serious damping due to anti-phase boundaries. Hence, the previous structural solution \cite{ksugiyama98_2by1} should be re-examined in view of that it could correspond to the average charge density map of the eight translates \{$\rho({\bf x}+{\bf s})$\} of the true charge density map $\rho({\bf x})$, where {\bf s} runs over the eight independent translation vectors ${\bf 0}$, ${\bf R}^\prime_1$, ${\bf R}^\prime_2$, ${\bf R}^\prime_3$, ${\bf R}^\prime_1+{\bf R}^\prime_2$, ${\bf R}^\prime_2+{\bf R}^\prime_3$,  ${\bf R}^\prime_3+{\bf R}^\prime_1$ and ${\bf R}^\prime_1+{\bf R}^\prime_2+{\bf R}^\prime_3$.

\begin{figure}
\caption{\label{fig:12}The `fake' cluster $\mathfrak{I}$ that is observed at the body center of the halved unit cell in the superposition of the eight translated versions of the true structure. Each of the shells has at least the point symmetry $m\bar{3}$ ($T_h$) when the contents of the sites are taken into consideration. Members of a shell with different contents are distinguished by colors: lg (light gray), dg (dark gray) and bk (black).}
\includegraphics[width=8cm]{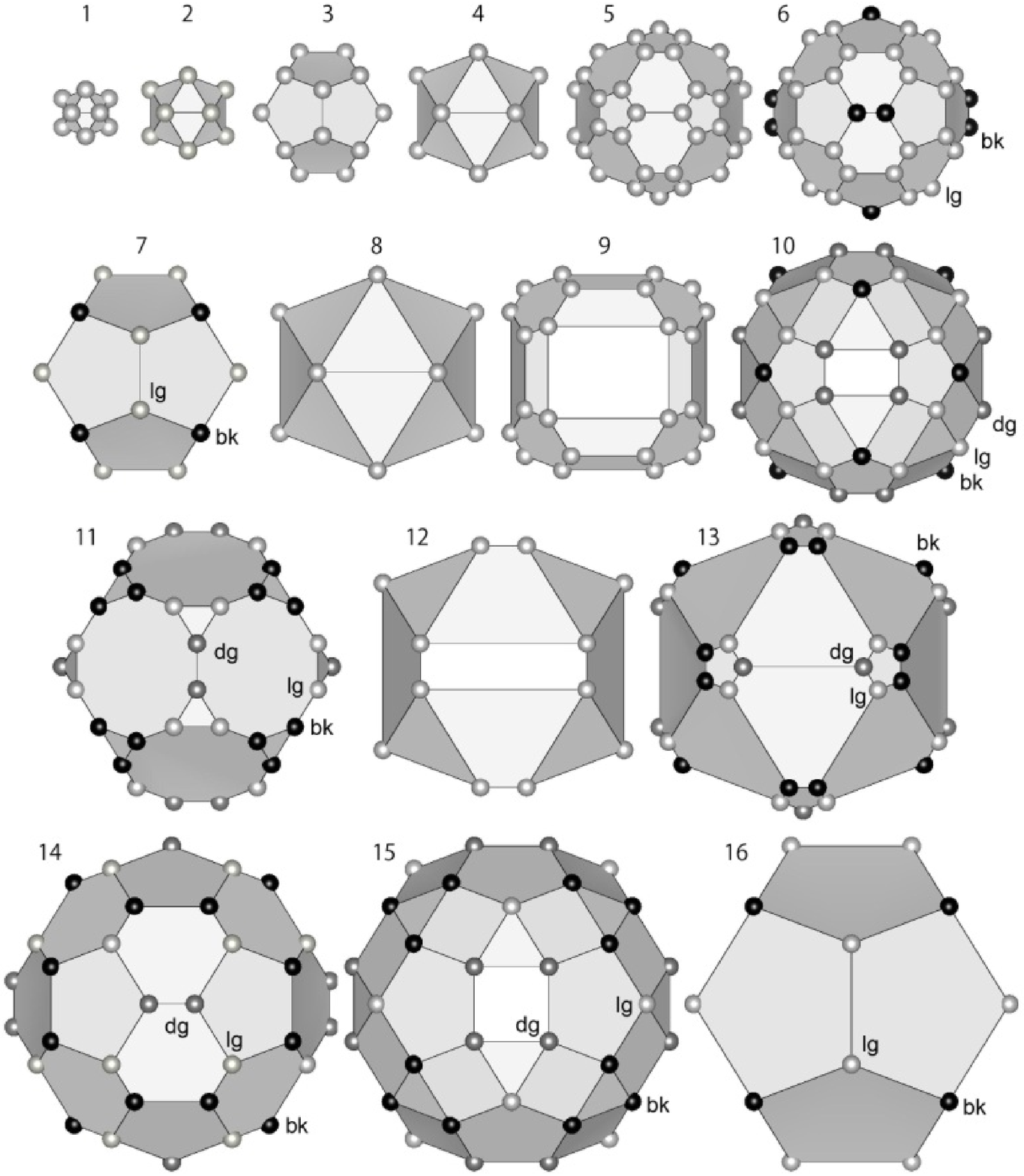}
\end{figure}

\begin{figure}
\caption{\label{fig:13}The `fake' cluster $\mathfrak{V}$ that is observed at the origin in the superposition of the eight translated versions of the true structure. Each of the shells has at least the point symmetry $m\bar{3}$ ($T_h$) when the contents of the sites are taken into consideration. Members of a shell with different contents are distinguished by colors: lg (light gray), dg (dark gray) and bk (black).}
\includegraphics[width=8cm]{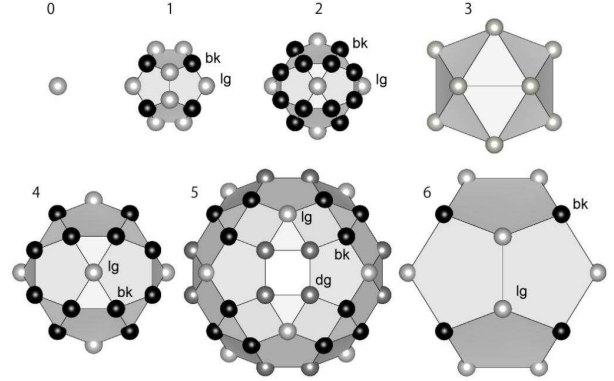}
\end{figure}

Let us take the superposition of the corresponding eight translates of the true crystal structure (Subsection \S\S\ref{subsec:packinggeometry}). Here, several atomic sites can fall onto the same position with a maximal degeneracy of 8. We find that cluster like arrangements are formed at the origin as well as at the body center of the half-sized unit cell for the superposed structure. Figure \ref{fig:12} illustrates the shell structure of the `fake' cluster $\mathfrak{I}$ at the body center, where most of the shells exhibit the icosahedral symmetry in the geometrical arrangement except the shells $\mathfrak{I}$-9 and $\mathfrak{I}$-12. This agrees remarkably well with the shell structures of the primary (i.e., giant) cluster reported by Sugiyama {\it et al.} \cite{ksugiyama98_2by1}. Another `fake' cluster $\mathfrak{V}$ with icosahedral arrangement is also found at the vertex of the unit cell; see Figure \ref{fig:13}. Importantly, the site degeneracies of a single shell in $\mathfrak{I}$ or $\mathfrak{V}$ generally break the icosahedral symmetry (Figures 12, 13 and Table \ref{table:siteinfakecluster}), where the minimal point group is $m\bar{3}$ ($T_h$). The violation of the icosahedral symmetry in $\mathfrak{V}$ may also account for the appearance of the secondary cluster reported by Sugiyama {\it et al.} \cite{ksugiyama98_2by1}. Therefore, the crystal structure reported by Sugiyama {\it et al.} \cite{ksugiyama98_2by1} is most likely to be an artifact in which the true structure is folded into half-sized unit cell.

\begin{table}
\caption{\label{table:siteinfakecluster}The content of each site in the fake clusters $\mathfrak{I}$ and $\mathfrak{V}$, which are depicted in Figures 12 and 13, respectively. The shells are numbered in increasing order of radius, where the central site of $\mathfrak{V}$ is numbered 0.}
\begin{center}
\begin{tabular}{lll}
 shell \# & radius& content \\
                &  [\AA]  & \\\\
  \multicolumn{3}{l}{(cluster $\mathfrak{I}$)} \\
 1 & 1.738 & 2$\langle$B$_5\rangle$ \\
 2 & 2.813 & 2$\langle$B$_3$M$_5\rangle$, 4$\langle$B$^2_3$M$_5\rangle$ \\
 3 & 4.144 &  2$\langle$M$^3_2\rangle$, 6$\langle$B$_5$M$^2_2\rangle$\\
 4 & 4.551 &  2$\langle$B$_0\rangle$\\
 5 & 5.550 &  2$\langle$B$_3$M$_3\rangle$\\
 6 & 6.330 &  lg: $\langle$B$_5$M$^2_2\rangle$, $\langle$B$_5$M$_2\rangle$, 4$\langle$M$_2\rangle$\\
              & &  bk: 2$\langle$B$_5$M$_2\rangle$, 4$\langle$M$_2\rangle$\\
 7 & 6.705 &  lg: 6$\langle$B$^2_3$M$^2_5\rangle$, 2$\langle$B$_3$M$^3_5\rangle$\\
              & &  bk: 3$\langle$B$^2_3$M$^2_5\rangle$, 5$\langle$B$_3$M$^3_5\rangle$\\
 8 & 7.364 &  2$\langle$B$_5$M$_2\rangle$, 6$\langle$M$_0\rangle$\\
 9 & 7.935 &  $\langle$M$_2\rangle$\\
 10 & 8.237 &  dg: $\langle$B$^2_3$M$_5\rangle$, $\langle$B$^2_3\rangle$, 6$\langle$B$_3$M$_3\rangle$\\
              & &  lg: $\langle$B$^2_3\rangle$, 5$\langle$B$_3$M$_3\rangle$\\
             & &  bk: 2$\langle$B$^2_3$M$_5\rangle$, 6$\langle$B$_3$M$_3\rangle$\\
 11 & 8.782 &  dg: 4$\langle$B$_5$M$^2_2\rangle$, 4$\langle$B$_5$M$_2\rangle$\\
             & &  lg: 6$\langle$B$_5$M$^2_2\rangle$, 2$\langle$B$_5$M$_2\rangle$\\
   & &  bk: 3$\langle$B$_5$M$^2_2\rangle$, $\langle$M$^3_2\rangle$, 2$\langle$B$_5$M$_2\rangle$, 2$\langle$M$_2\rangle$\\
 12 & 9.526 &  2$\langle$B$_3$M$_3\rangle$\\
 13 & 10.001 &  dg: 2$\langle$M$^2_2\rangle$, 4$\langle$M$_2\rangle$\\
              & &  lg: $\langle$M$_2\rangle$\\
              & &  bk: $\langle$M$^2_2\rangle$, 3$\langle$M$_2\rangle$\\
 14 & 10.242 &  dg: 4$\langle$B$^2_3$M$_5\rangle$, 4$\langle$B$^2_3$M$^2_5\rangle$\\
              & &  lg: $\langle$B$_3$M$_3\rangle$, $\langle$B$^2_3\rangle$, 4$\langle$B$^2_3$M$_5\rangle$, 2$\langle$B$^2_3$M$^2_5\rangle$\\
              & &  bk: $\langle$B$_3$M$_3\rangle$, 3$\langle$B$^2_3$M$_5\rangle$, 4$\langle$B$^2_3$M$^2_5\rangle$\\
 15 & 10.685 &  dg: 2$\langle$B$_5$M$_2\rangle$, 6$\langle$B$_5$M$^2_2\rangle$\\
              & &  lg: 8$\langle$B$_5$M$^2_2\rangle$\\
              & &  bk: $\langle$M$_0\rangle$, $\langle$B$_5$M$_2\rangle$, 6$\langle$B$_5$M$^2_2\rangle$\\
 16 & 10.849 &  lg: 8$\langle$B$_0\rangle$\\
              & &  bk: 5$\langle$B$_0\rangle$\\ \\
 \multicolumn{3}{l}{(cluster $\mathfrak{V}$)} \\
 0 & 0.000 &  8$\langle$M$_0\rangle$\\
 1 & 2.561 &  lg: 2$\langle$B$_3$M$_3\rangle$\\
              & &  bk: 5$\langle$B$_3$M$_3\rangle$\\
 2 & 2.957 &  lg: 4$\langle$M$_2\rangle$\\
              & &  bk: $\langle$M$_2\rangle$\\
 3 & 4.551 &  8$\langle$B$^2_3$M$^2_5\rangle$\\
 4 & 4.785 &  lg: 4$\langle$B$_5$M$^2_2\rangle$, 4$\langle$M$_2\rangle$\\
             & &  bk: 7$\langle$B$_5$M$^2_2\rangle$, $\langle$M$_2\rangle$\\
 5 & 6.604 &  dg: $\langle$B$^2_3$M$_5\rangle$, 5$\langle$B$_3$M$_3\rangle$\\
              & &  lg: 4$\langle$B$^2_3$M$_5\rangle$, 2$\langle$B$_3$M$_3\rangle$\\
              & &  bk: $\langle$B$^2_3$M$_5\rangle$, 2$\langle$B$_3$M$_3\rangle$\\
 6 & 6.705 &  lg: 2$\langle$B$_0\rangle$\\
              & &  bk: 5$\langle$B$_0\rangle$\\
\end{tabular}
\end{center}
\end{table}

\subsection{Atomic packing}\label{subsec:atomicpacking}

As described in Subsection \S\S\ref{subsec:packinggeometry}, the closest interatomic distances in the idealized construction are given by $b^\prime$ and $c^\prime$ and are parallel to two- and three-fold axes, respectively. These distances are $1/\tau^2$ times smaller than b- and c-linkages for the cluster packing, so that the ratio between the two satisfies $c^\prime/b^\prime=c/b$ ($=\sqrt{3}/2$). Then, it is natural that atoms would be locally arranged to sit on the vertices of the four basic shapes, A$^\prime$, B$^\prime$, C$^\prime$ and D$^\prime$, which are the miniature versions (scale, $\times1/\tau^2$) of A-, B-, C- and D-cells, respectively. Yet, detailed inspection reveals several essential differences between the arrangement of atoms and that of clusters; that is, the former is not precisely a miniaturized version of the latter.

\begin{figure}
\caption{\label{fig:14} The two basic tiles, I$^\prime$ (left) and H$^\prime$ (right), that are necessary to provide a tiling description for the atomic positions. The double bars represent the b$^\prime$-linkages, while the thick bars the c$^\prime$-linkages between atoms. There is an atomic site for Pd at the center of the tile I$^\prime$. The surfaces of the tiles are composed of X$^\prime$- and Y$^\prime$-faces, which are the miniature versions of the X- and Y-faces, respectively. $+$ or $-$ signs shown on the front faces indicate which of the symmetrically distinguishable sides are facing outward. Note that the point symmetry of I$^\prime$ and H$^\prime$ are $\bar{5}\bar{3}2/m$ ($I_{\rm h}$) and $mmm$ ($D_{2h}$), respectively.}
\includegraphics[width=8cm]{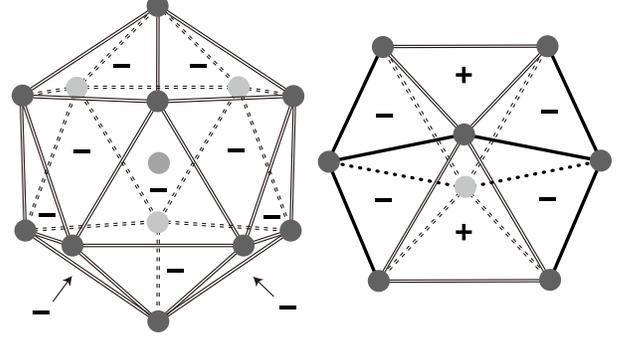}
\end{figure}

An important feature of the atomic packing is that there exists an icosahedral configuration located at the center of each B-cluster, consisting of the central site (type, $\langle$B$_0\rangle$) and the 12 surrounding sites (type, $\langle$B$_5$,...$\rangle$) forming the inner icosahedral shell. Hence, it is required to introduce an icosahedron with a central dot as a basic shape if a tiling description of the atomic packing is sought. We denote this shape I$^\prime$ (Figure \ref{fig:14}, left).

Remember also that there exist two splitting sites (type, $\langle$M$_2\rangle$) associated with each type I b-linkage that connects two M-clusters. Although the two sites cannot be occupied simultaneously, an Al atom is most likely jumping back and forth between these sites at a typical annealing temperature of $850^\circ{\rm C}$, where both the splitting sites play an inevitable role in maintaining and stabilizing the structure. The two splitting sites are always located at the two tips of a flat hexagonal bipyramid, which corresponds to the overlap of two adjacent M-clusters. We denote this shape H$^\prime$ (Figure \ref{fig:14}, right). The two tips of H$^\prime$ are the very splitting sites of the type $\langle$M$_2\rangle$, while the six vertices of the base hexagon include two atomic sites that are symbolized as $\langle$M$_5$,M$_5$,...$\rangle$ and four that are symbolized as $\langle$M$_2$,M$_2$,...$\rangle$.

\begin{figure}
\caption{\label{fig:15}Two kinds of voids associated with D-cells. (a) A void within the outer shell of an M-cluster located at a corner $D_a$ of a type I D-cell. The edges of the D-cell are represented as black bars. The spheres are the atomic positions predicted by the present geometrical rules, where labels for some atomic sites are presented in the figure. (b) Similar to (a) but the void is associated with a corner $D_b$ of a type II D-cell. A splitting position for an atomic site is indicated with an arrow and a dotted circle  (labeled `s'). A possible atomic displacement is indicated with an arrow (labeled `d'). In the refinement, the two kinds of splitting positions as in (a) and (b) are included as Pd157$^\prime$ and Al86$^\prime$, respectively.}
\includegraphics[width=8cm]{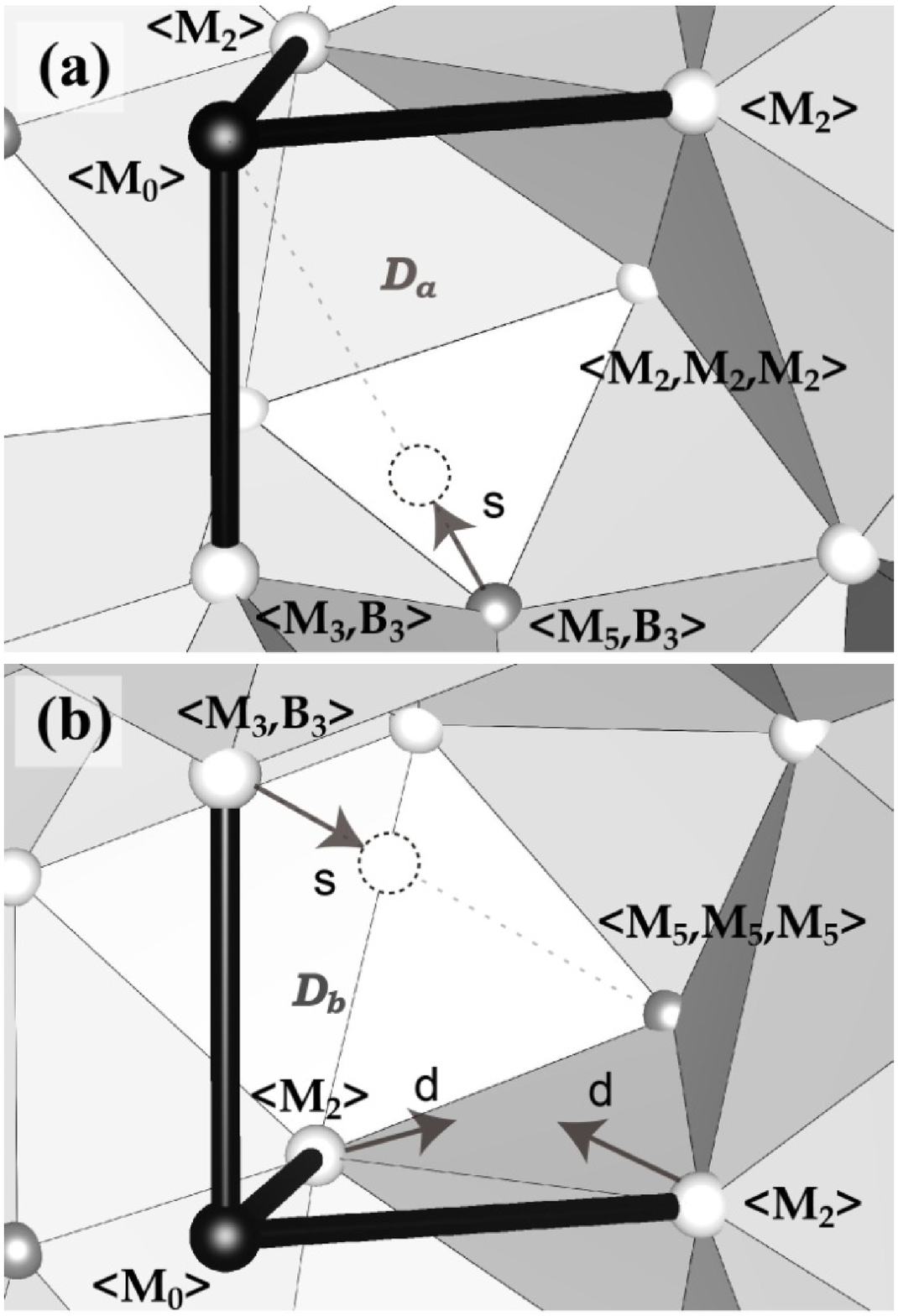}
\end{figure}

The idealized atomic arrangement can now be represented as an arrangement of the six basic geometrical shapes A$^\prime$, B$^\prime$, C$^\prime$, D$^\prime$, I$^\prime$ and H$^\prime$. Still, it may not be called a tiling because a part of the space is left blank in this description. The latter circumstance arises when the cluster packing involves any number of D-cells. Remember that a D-cell has the largest volume among the four canonical cells and clusters are free from interpenetration at its central region. This causes that the inner shell atoms of an M-cluster located at one of the vertices are sparsely distributed toward the center of the D-cell, thereby providing a large unfilled space or a void inside the outer shell of the M-cluster. Two relevant kinds of void are illustrated in Figure \ref{fig:15} for the type I and type II D-cells. These voids can be interconnected to form a larger void or a channel if D-cells adjoin together. In the refinement of the present approximant, it is observed that a surrounding atom of such a void can have a plitting position within the void (Pd157$^\prime$ for Figure \ref{fig:15}(a) and Al86$^\prime$ for Figure \ref{fig:15}(b)) or exhibit a significant displacement toward the center of the void (Al174$^\prime$ for Figure \ref{fig:15}(b)). These facts are physically consistent with the present geometrical description.

\section{Concluding remarks}\label{sec6}

A simple heuristic search in the quaternary Al-Pd-Cr-Fe system has led to the first discovery of a highly stable cubic approximant phase to Al-based F-type icosahedral quasicrystals $i$-Al$_{70}$Pd$_{20}$TM$_{10}$ (TM=transition metals). Single crystals of the approximant a few hundred $\mu m$'s in diameter have been grown with slow cooling, suggesting that the approximant phase is formed congruently from the melt. Using single crystal X-ray diffraction, an ab initio structure analysis has been carried out successfully, revealing a crystal symmetry of $Pa\bar{3}$ with a lattice constant of 40.5 \AA. All the atoms in the crystal belong to two kinds of clusters which in the literature are referred to as the Mackay type and the Bergman type clusters, whereas there is no need for glue atoms.

The centers of the clusters are located at the vertices of a CCT, designated as a 2$\times$2$\times$2 superstructure of the 3/2-packing \cite{clhenley91}, in which the parity of each vertex determines unambiguously the kind of the relevant cluster. Importantly, the present structure offers a set of universal rules for the local packing of clusters. The rules are applicable to arbitrary CCT's, whereby a number of hypothetical approximants can be constructed. Some of these hypothetical structures could be synthesized in related alloys. Indeed, the crystal structure that was previously reported as a cubic 1/1 approximant  by Sugiyama {\it et al.} \cite{ksugiyama98_1by1} could be better understood as an artifact of folding a hypothetical structure constructed from a 2$\times$2$\times$2 superstructure of the 2/1 packing \cite{clhenley91} into a half-sized unit cell. Here a similar argument to that presented in Subsection \S\S\ref{subsec:antiphaseboundaries} is assumed. A search for stable approximants with various degrees of rational approximation to $\tau$ within related alloy systems is an important future challenge.

In actual fact, the proposed two cluster units, M and B, have been already used (with a somewhat different way of handling the inner shell of M-cluster) to compose a hypothetical model of F-type icosahedral quasicrystals, e.g., {\it i}-Al-Pd-Mn \cite{velser96,gratiasreview99}. In this case, the clusters are assumed to be centered at the vertices of a random tiling composed of two kinds of rhombohedra called Ammann's rhombohedra. Accordingly individual clusters are connected with their adjacent neighbors through the edges (in five-fold directions) or the short diagonals (in two-fold directions) of the rhombic faces, whereas clusters at the two ends of the short body diagonal (along a three-fold direction) of each obtuse rhombohedron are forced to be interpenetrated heavily with each other. Our findings proved to be against this model, since no pair of clusters connected through a five-fold linkage is observed in the reconstructed crystal structure.

The superlattice ordering in the icosahedral quasicrystal, $i$-Al-Pd-Mn, \cite{ishimasa92} has once attracted much attention, and rich features associated with this phenomenon have been published \cite{boissieu98,audier99,letoublon00,hirai00}. However, no clear account on its physical origin has been provided, although it seems to be closely related to the binary cluster composition for the F-type icosahedral systems. In the expectation that the superlattice ordering in the present approximant may be related to that in the quasicrystal, possible P-type superlattice orderings in the six-dimensional F-type hypercubic lattice have been analyzed in Subsection \S\S\ref{subsec:phason}, where it has turned out that the two phenomena cannot be associated with the same six-dimensional superlattice. Perhaps, a kind of lock-in transition facilitated by the discreteness of the atomic surfaces is more likely to be the origin of the superlattice ordering in the approximant, whereas in the quasicrystal more subtle competitions between the boundaries of atomic surfaces could be responsible for it.


One of the most important outcomes of this work is the finding that the atomic arrangement is perfectly described using a set of geometrical rules for cluster packing defined on a CCT. Local arrangements of atomic positions thus determined can be described generally as the vertices of six basic polyhedra A$^\prime$, B$^\prime$, C$^\prime$, D$^\prime$, I$^\prime$ and H$^\prime$, which are arranged under the face-to-face matching constraint. Importantly, the latter construction leaves the presence of a void inside the outer shell of each M-cluster that is located at a vertex of a D-cell. The present structure refinement has indicated that some of the surrounding atoms of the voids have a secondary (i.e., splitting) position within the void or exhibit a significant displacement toward the center of the void.

Altogether, the present results have greatly improved our understandings on the local atomic structures of F-type icosahedral quasicrystals and their approximants. Clearly, the present local rules for cluster packing maintain a certain level of universality, and a variety of complex structures that are closely related to Al-based F-type icosahedral quasicrystals are expected to be described within the same framework. However, when the structure of the quasicrystal is at stake, there remains a crucial issue of whether the canonical cells can be used to construct a quasiperiodic tiling with the icosahedral symmetry. The peculiar shapes of the canonical cells are known to impose non-local geometrical constraints, making it difficult to determine their non-periodic alignments throughout the whole space. It is an important challenge in the theoretical crystallography of quasicrystals to prove if an icosahedral quasiperiodic tiling can be constructed using the canonical cells.

\appendix

\section{Derivation of the statistics formulae for the 16 types of atomic sites}\label{appendix:sitestat}

Note that each class of atomic sites is associated with one or more type(s) of geometric objects (see Figure \ref{fig:8} and Table \ref{table:sitesymbols}), and a careful inspection of the decorated objects proves the following relationships:
\begin{eqnarray}
n({\rm node_I}) &=& f \\
&=& (h+m)/20 \\
&=& (n+3r+s+2t+3u)/12\\
&=& (g+2k+l+2p+3q)/30 ,\\
n({\rm node_{II}}) &=& i \\
&=& (j+l+p)/12 \\
&=& (m+n+2o+2s+2t+u)/20 ,\\
n({\rm b_I}) &=& g/2 \\
&=& (3r+t+3u)/2 \\
&=& (k+p+3q)/4 ,\\
n({\rm b_{II}}) &=& (o+s+t)/2 ,\\
n({\rm c}) &=& m \\
&=& (l+2p)/3 \\
&=& (n+2s+4t+3u)/3 ,\\
n({\rm X_I}) &=& p \\
&=& 2t+3u,\\
n({\rm X_{II}}) &=& s+2t,\\
n({\rm Y_I}) &=& q \\
&=& r+u,\\
n({\rm A_I}) &=& t,\\
n({\rm C_I}) &=& u,
\end{eqnarray}
where the variables $f$, $g$, ..., $u$ represent the number frequencies for the 16 types of atomic sites; see Table \ref{table:sitesymbols} for the definitions. Some of these equations turn out to be redundant in the sense they are not necessarily independent with each other. By solving these equations, Eqs.(\ref{eq_sitestat_i}) - (\ref{eq_sitestat_f}) of Subsection \S\S\ref{subsec:canonicalcells} can be readily obtained.

\section{Refined parameters}\label{appendixB}

Among the 4728 atomic positions contained in a single unit cell, 4680 are reproduced (within small deviations) by replicating the cluster templates at the vertices of the CCT under the geometrical rules fully described in Subsection \S\S\ref{subsec:packinggeometry}. The remaining 48 positions are irregular splitting sites (Al86$^\prime$ and Pd157$^\prime$) introduced inside voids associated with M-clusters centered at Fe/Pd82 and Cr/Al98. The inclusion of these splitting sites has reduced $R_w(F)$ by more than 1\%. Bear in mind that no glue atom is included in the structure.

The existence of voids causes some additional irregularities for the refinement: (1) The splitting site Pd157$^\prime$ appears to contradict with the site Al/Cr104 which lies nearby. Therefore, the total s.o.f. of Al/Cr104 is set to be equal to that of Pd157, so that it complements that of Pd157$^\prime$. (2) In addition, the atomic site Al174$^\prime$ shifted toward the center of a void lies too close to Al183, whereas the distance between Al174 and Al174$^\prime$, which are originally splitting sites for a single Al atom, is large enough to accommodate atoms at both of them. Therefore, s.o.f.'s for the four positions Al174, Al174$^\prime$, Al183, Al183$^\prime$ are refined so that their sum is fixed to 2. These are exceptional measured that we have taken to achieve reasonable convergence of the refinement.

The number of independent atomic positions cannot be reduced below 204. As an effort to suppress the number of independent parameters, the following three constraints on the isotropic atomic displacement parameters (IADP's) are introduced. (1) Within the inner shell of each B-cluster, the IADP's are set to be identical for the same site type. See, for example, the IADP's of Al12, ..., Al23 in Table \ref{table:AtomicPara}; the IADP's for the site classes $\langle$B$_5$,M$_2$,M$_2\rangle$,  $\langle$B$_5$,M$_2\rangle$ and  $\langle$B$_5\rangle$ are 0.015, 0.01 and 0.006, respectively. (2) The IADP's for the two splitting sites associated with each b-linkage between M-clusters are set to be identical; e.g., the IADP's for Al173 and Al173$^\prime$ are both 0.029. (3) In the final stage of the refinement, it was found that the IADP and the site occupation factor (s.o.f.) are strongly correlated for Fe/Pd154. If both the parameters are refined, the IADP converges to a negative value while the Pd concentration reduces by more than 10 \%. Hence, the IADP of this atomic site is fixed to be 0.011.

In Table \ref{table:AtomicPara}, it is observed that IADP's for some atomic positions in the inner shell of M-clusters are exceptionally large (e.g., Al96, Al97, Al00 and Al/Cr104). This might be a due to possible disorders in the M$_3$ shells. A full characterization of such disorders may require a detailed analysis of charge density distributions for the M$_3$ shells. However, the quality of our intensity data may not be sufficient for performing such an analysis.

\setlength{\tabcolsep}{1mm}
\begin{longtable*}{lcl lll lll}
\caption{Table for atomic parameters for the refined structure. $s.o.f.$ gives the site occupation factor for each atomic species. $U$ is the IADP. $\Delta$ gives the distance between the refined atomic position and the ideal position.}\label{table:AtomicPara}\\
\hline\\
atom & {\tiny Wyckoff} & $s.o.f.$ & $x/a_{lat}$ & $y/a_{lat}$ & $z/a_{lat}$ & $U$ & site type & $\Delta/a_{lat}$\\ \hline
\endfirsthead
\hline\\
atom & {\tiny Wyckoff} & $s.o.f.$ & $x/a_{lat}$ & $y/a_{lat}$ & $z/a_{lat}$ & $U$ & site type & $\Delta/a_{lat}$\\ \hline
\multicolumn{9}{r}{(continued from the previous page)}
\endhead
\multicolumn{9}{r}{(continue to next page)}\\ \hline
\endfoot
\multicolumn{9}{r}{(end of table)}\\
\hline\\
\endlastfoot
Pd1 & $8c$ &  1.00 & 0.09485(8) & 0.09485(8) & 0.09485(8) &  0.003 & $\langle$B$_0^{(1)}$$\rangle$ &  0.001\\
Al2 & $24d$ &  1.00 & 0.1308(3) & 0.0944(3) & 0.0381(3) &  0.005 & $\langle$B$_5^{(1)}$,M$_2^{(3)}$$\rangle$ &  0.002\\
Al3 & $24d$ &  1.00 & 0.0585(4) & 0.0946(4) & 0.0386(3) &  0.005 & $\langle$B$_5^{(1)}$,M$_2^{(1)}$,M$_2^{(3)}$$\rangle$ &  0.002\\
Al4 & $24d$ &  1.00 & 0.1516(4) & 0.1300(3) & 0.0949(4) &  0.005 & $\langle$B$_5^{(1)}$,M$_2^{(5)}$,M$_2^{(5)}$$\rangle$ &  0.004\\
Al5 & $24d$ &  1.00 & 0.0948(4) & 0.1515(4) & 0.0597(4) &  0.005 & $\langle$B$_5^{(1)}$,M$_2^{(3)}$,M$_2^{(5)}$$\rangle$ &  0.003\\
Pd6 & $8c$ &  1.00 & 0.40400(9) & 0.40400(9) & 0.40400(9) &  0.008 & $\langle$B$_0^{(2)}$$\rangle$ &  0.001\\
Al7 & $24d$ &  1.00 & 0.4612(4) & 0.4387(4) & 0.4051(4) &  0.012 & $\langle$B$_5^{(2)}$,M$_2^{(2)}$,M$_2^{(4)}$$\rangle$ &  0.003\\
Al8 & $24d$ &  1.00 & 0.4043(4) & 0.4619(3) & 0.3680(3) &  0.002 & $\langle$B$_5^{(2)}$,M$_2^{(4)}$$\rangle$ &  0.002\\
Al9 & $24d$ &  1.00 & 0.3698(4) & 0.4025(4) & 0.3458(4) &  0.012 & $\langle$B$_5^{(2)}$,M$_2^{(7)}$,M$_2^{(7)}$$\rangle$ &  0.003\\
Al10 & $24d$ &  1.00 & 0.4389(4) & 0.4046(4) & 0.3487(4) &  0.012 & $\langle$B$_5^{(2)}$,M$_2^{(4)}$,M$_2^{(7)}$$\rangle$ &  0.004\\
Pd11 & $24d$ &  1.00 & 0.25036(9) & 0.3449(1) & 0.19066(9) &  0.006 & $\langle$B$_0^{(3)}$$\rangle$ &  0.001\\
Al12 & $24d$ &  1.00 & 0.2868(4) & 0.3453(4) & 0.1326(4) &  0.015 & $\langle$B$_5^{(3)}$,M$_2^{(3)}$,M$_2^{(6)}$$\rangle$ &  0.001\\
Al13 & $24d$ &  1.00 & 0.3459(4) & 0.2472(4) & 0.2142(5) &  0.015 & $\langle$B$_5^{(3)}$,M$_2^{(6)}$,M$_2^{(7)}$$\rangle$ &  0.003\\
Al14 & $24d$ &  1.00 & 0.2503(3) & 0.4026(4) & 0.2255(4) &  0.010 & $\langle$B$_5^{(3)}$,M$_2^{(7)}$$\rangle$ &  0.003\\
Al15 & $24d$ &  1.00 & 0.1915(4) & 0.3809(4) & 0.1905(4) &  0.010 & $\langle$B$_5^{(3)}$,M$_2^{(6)}$$\rangle$ &  0.001\\
Al16 & $24d$ &  1.00 & 0.3074(4) & 0.3101(4) & 0.1894(4) &  0.015 & $\langle$B$_5^{(3)}$,M$_2^{(6)}$,M$_2^{(7)}$$\rangle$ &  0.003\\
Al17 & $24d$ &  1.00 & 0.2508(3) & 0.2898(4) & 0.2264(3) &  0.006 & $\langle$B$_5^{(3)}$$\rangle$ &  0.004\\
Al18 & $24d$ &  1.00 & 0.3074(4) & 0.3804(4) & 0.1899(4) &  0.015 & $\langle$B$_5^{(3)}$,M$_2^{(3)}$,M$_2^{(7)}$$\rangle$ &  0.002\\
Al19 & $24d$ &  1.00 & 0.2160(4) & 0.3474(4) & 0.1333(4) &  0.010 & $\langle$B$_5^{(3)}$,M$_2^{(5)}$$\rangle$ &  0.003\\
Al20 & $24d$ &  1.00 & 0.2490(4) & 0.4013(4) & 0.1564(4) &  0.010 & $\langle$B$_5^{(3)}$,M$_2^{(3)}$$\rangle$ &  0.004\\
Al21 & $24d$ &  1.00 & 0.2861(4) & 0.3468(4) & 0.2471(4) &  0.015 & $\langle$B$_5^{(3)}$,M$_2^{(7)}$,M$_2^{(7)}$$\rangle$ &  0.003\\
Al22 & $24d$ &  1.00 & 0.1935(4) & 0.3102(4) & 0.1902(4) &  0.015 & $\langle$B$_5^{(3)}$,M$_2^{(5)}$,M$_2^{(6)}$$\rangle$ &  0.003\\
Al23 & $24d$ &  1.00 & 0.2492(4) & 0.2876(5) & 0.1557(4) &  0.015 & $\langle$B$_5^{(3)}$,M$_2^{(5)}$,M$_2^{(6)}$$\rangle$ &  0.002\\
Pd24 & $24d$ &  1.00 & 0.34558(9) & 0.49947(11) & 0.2496(1) &  0.008 & $\langle$B$_0^{(4)}$$\rangle$ &  0.001\\
Al25 & $24d$ &  1.00 & 0.2514(4) & 0.2123(4) & 0.0366(4) &  0.010 & $\langle$B$_5^{(4)}$,M$_2^{(5)}$,M$_2^{(6)}$$\rangle$ &  0.002\\
Al26 & $24d$ &  1.00 & 0.4019(5) & 0.4640(4) & 0.2491(4) &  0.010 & $\langle$B$_5^{(4)}$,M$_2^{(4)}$,M$_2^{(7)}$$\rangle$ &  0.003\\
Al27 & $24d$ &  1.00 & 0.3793(4) & 0.4985(4) & 0.3065(4) &  0.012 & $\langle$B$_5^{(4)}$,M$_2^{(4)}$$\rangle$ &  0.004\\
Al28 & $24d$ &  1.00 & 0.2881(4) & 0.4654(4) & 0.2491(4) &  0.012 & $\langle$B$_5^{(4)}$,M$_2^{(7)}$$\rangle$ &  0.003\\
Al29 & $24d$ &  1.00 & 0.1930(4) & 0.1181(4) & -0.0004(4) &  0.012 & $\langle$B$_5^{(4)}$,M$_2^{(3)}$$\rangle$ &  0.002\\
Al30 & $24d$ &  1.00 & 0.3461(4) & 0.4437(4) & 0.2840(4) &  0.010 & $\langle$B$_5^{(4)}$,M$_2^{(7)}$,M$_2^{(7)}$$\rangle$ &  0.004\\
Al31 & $24d$ &  1.00 & 0.3457(4) & 0.4426(4) & 0.2158(4) &  0.010 & $\langle$B$_5^{(4)}$,M$_2^{(3)}$,M$_2^{(7)}$$\rangle$ &  0.003\\
Al32 & $24d$ &  1.00 & 0.2834(4) & 0.1540(4) & 0.0569(4) &  0.010 & $\langle$B$_5^{(4)}$,M$_2^{(5)}$,M$_2^{(6)}$$\rangle$ &  0.004\\
Al33 & $24d$ &  1.00 & 0.2132(4) & 0.1562(4) & 0.0572(4) &  0.010 & $\langle$B$_5^{(4)}$,M$_2^{(5)}$,M$_2^{(5)}$$\rangle$ &  0.003\\
Al34 & $24d$ &  1.00 & 0.1921(4) & 0.1897(4) & 0.0007(4) &  0.010 & $\langle$B$_5^{(4)}$,M$_2^{(3)}$,M$_2^{(5)}$$\rangle$ &  0.002\\
Al35 & $24d$ &  1.00 & 0.3056(4) & 0.1904(4) & -0.0001(4) &  0.010 & $\langle$B$_5^{(4)}$,M$_2^{(6)}$,M$_2^{(7)}$$\rangle$ &  0.003\\
Al36 & $24d$ &  1.00 & 0.2489(4) & 0.0980(5) & 0.0352(4) &  0.010 & $\langle$B$_5^{(4)}$,M$_2^{(4)}$,M$_2^{(5)}$$\rangle$ &  0.003\\
Pd37 & $24d$ &  1.00 & 0.40403(9) & 0.09507(9) & 0.09482(9) &  0.006 & $\langle$B$_0^{(5)}$$\rangle$ &  0.001\\
Al38 & $24d$ &  1.00 & 0.4391(4) & 0.0940(4) & 0.0384(4) &  0.007 & $\langle$B$_5^{(5)}$,M$_2^{(2)}$,M$_2^{(4)}$$\rangle$ &  0.003\\
Al39 & $24d$ &  1.00 & 0.1299(4) & 0.4040(4) & 0.0383(4) &  0.008 & $\langle$B$_5^{(5)}$,M$_2^{(4)}$$\rangle$ &  0.003\\
Al40 & $24d$ &  1.00 & 0.3692(4) & 0.0969(4) & 0.0380(3) &  0.008 & $\langle$B$_5^{(5)}$,M$_2^{(4)}$$\rangle$ &  0.002\\
Al41 & $24d$ &  1.00 & 0.0595(4) & 0.4051(4) & 0.0375(4) &  0.007 & $\langle$B$_5^{(5)}$,M$_2^{(2)}$,M$_2^{(4)}$$\rangle$ &  0.001\\
Al42 & $24d$ &  1.00 & 0.0944(4) & 0.4614(4) & 0.0600(4) &  0.007 & $\langle$B$_5^{(5)}$,M$_2^{(2)}$,M$_2^{(4)}$$\rangle$ &  0.003\\
Al43 & $24d$ &  1.00 & 0.4616(4) & 0.1302(4) & 0.0957(4) &  0.008 & $\langle$B$_5^{(5)}$,M$_2^{(4)}$$\rangle$ &  0.003\\
Al44 & $24d$ &  1.00 & 0.4043(4) & 0.1518(4) & 0.0607(4) &  0.007 & $\langle$B$_5^{(5)}$,M$_2^{(4)}$,M$_2^{(6)}$$\rangle$ &  0.003\\
Al45 & $24d$ &  1.00 & 0.1523(4) & 0.3699(4) & 0.0951(4) &  0.008 & $\langle$B$_5^{(5)}$,M$_2^{(5)}$$\rangle$ &  0.003\\
Al46 & $24d$ &  1.00 & 0.3477(4) & 0.1309(4) & 0.0954(4) &  0.007 & $\langle$B$_5^{(5)}$,M$_2^{(5)}$,M$_2^{(6)}$$\rangle$ &  0.002\\
Al47 & $24d$ &  1.00 & 0.4047(4) & 0.1521(4) & 0.1314(4) &  0.008 & $\langle$B$_5^{(5)}$,M$_2^{(6)}$$\rangle$ &  0.002\\
Al48 & $24d$ &  1.00 & 0.1520(4) & 0.4391(4) & 0.0946(4) &  0.008 & $\langle$B$_5^{(5)}$,M$_2^{(4)}$$\rangle$ &  0.003\\
Al49 & $24d$ &  1.00 & 0.0951(4) & 0.3472(4) & 0.0605(4) &  0.007 & $\langle$B$_5^{(5)}$,M$_2^{(4)}$,M$_2^{(5)}$$\rangle$ &  0.002\\
Pd50 & $24d$ &  1.00 & 0.5000(1) & 0.25064(9) & 0.15386(9) &  0.004 & $\langle$B$_0^{(6)}$$\rangle$ &  0.001\\
Al51 & $24d$ &  1.00 & 0.3062(4) & 0.3107(4) & 0.0003(4) &  0.009 & $\langle$B$_5^{(6)}$,M$_2^{(6)}$,M$_2^{(7)}$$\rangle$ &  0.003\\
Al52 & $24d$ &  1.00 & 0.4989(4) & 0.1934(4) & 0.1882(4) &  0.009 & $\langle$B$_5^{(6)}$,M$_2^{(5)}$$\rangle$ &  0.004\\
Al53 & $24d$ &  1.00 & 0.4638(4) & 0.2495(4) & 0.0970(4) &  0.009 & $\langle$B$_5^{(6)}$,M$_2^{(4)}$,M$_2^{(6)}$$\rangle$ &  0.002\\
Al54 & $24d$ &  1.00 & 0.4985(4) & 0.3065(4) & 0.1190(4) &  0.009 & $\langle$B$_5^{(6)}$,M$_2^{(3)}$$\rangle$ &  0.003\\
Al55 & $24d$ &  1.00 & 0.5359(4) & 0.2490(4) & 0.2103(4) &  0.009 & $\langle$B$_5^{(6)}$,M$_2^{(5)}$,M$_2^{(6)}$$\rangle$ &  0.003\\
Al56 & $24d$ &  1.00 & 0.4998(4) & 0.1930(4) & 0.1203(4) &  0.009 & $\langle$B$_5^{(6)}$,M$_2^{(4)}$$\rangle$ &  0.003\\
Al57 & $24d$ &  1.00 & 0.4436(4) & 0.2836(4) & 0.1542(4) &  0.009 & $\langle$B$_5^{(6)}$,M$_2^{(6)}$,M$_2^{(7)}$$\rangle$ &  0.004\\
Al58 & $24d$ &  1.00 & 0.4417(4) & 0.2141(4) & 0.1543(4) &  0.009 & $\langle$B$_5^{(6)}$,M$_2^{(6)}$$\rangle$ &  0.001\\
Al59 & $24d$ &  1.00 & 0.2151(4) & 0.3465(4) & 0.0582(4) &  0.009 & $\langle$B$_5^{(6)}$,M$_2^{(5)}$$\rangle$ &  0.002\\
Al60 & $24d$ &  1.00 & 0.2841(4) & 0.3448(4) & 0.0568(4) &  0.009 & $\langle$B$_5^{(6)}$,M$_2^{(3)}$,M$_2^{(6)}$$\rangle$ &  0.003\\
Al61 & $24d$ &  1.00 & 0.2502(4) & 0.4030(4) & 0.0342(4) &  0.009 & $\langle$B$_5^{(6)}$,M$_2^{(3)}$,M$_2^{(4)}$$\rangle$ &  0.003\\
Al62 & $24d$ &  1.00 & 0.4650(4) & 0.2496(4) & 0.2105(4) &  0.009 & $\langle$B$_5^{(6)}$,M$_2^{(7)}$$\rangle$ &  0.003\\
Pd63 & $24d$ &  1.00 & 0.49994(9) & 0.44058(9) & 0.15447(8) &  0.007 & $\langle$B$_0^{(7)}$$\rangle$ &  0.000\\
Al64 & $24d$ &  1.00 & 0.3822(4) & 0.3091(4) & 0.0010(4) &  0.012 & $\langle$B$_5^{(7)}$,M$_2^{(6)}$,M$_2^{(7)}$$\rangle$ &  0.001\\
Al65 & $24d$ &  1.00 & 0.4633(4) & 0.4390(4) & 0.2120(4) &  0.012 & $\langle$B$_5^{(7)}$,M$_2^{(4)}$,M$_2^{(7)}$$\rangle$ &  0.003\\
Al66 & $24d$ &  1.00 & 0.5340(4) & 0.4421(4) & 0.2129(4) &  0.012 & $\langle$B$_5^{(7)}$,M$_2^{(4)}$,M$_2^{(6)}$$\rangle$ &  0.003\\
Al67 & $24d$ &  1.00 & 0.4425(4) & 0.4024(4) & 0.0353(4) &  0.012 & $\langle$B$_5^{(7)}$,M$_2^{(1)}$,M$_2^{(3)}$$\rangle$ &  0.003\\
Al68 & $24d$ &  1.00 & 0.4756(3) & 0.3449(3) & 0.0579(4) &  0.008 & $\langle$B$_5^{(7)}$,M$_2^{(3)}$$\rangle$ &  0.002\\
Al69 & $24d$ &  1.00 & 0.4431(4) & 0.4768(3) & 0.1556(3) &  0.008 & $\langle$B$_5^{(7)}$,M$_2^{(3)}$$\rangle$ &  0.002\\
Al70 & $24d$ &  1.00 & 0.4986(4) & 0.3836(3) & 0.1198(3) &  0.008 & $\langle$B$_5^{(7)}$,M$_2^{(3)}$$\rangle$ &  0.003\\
Al71 & $24d$ &  1.00 & 0.5003(4) & 0.4976(4) & 0.1184(3) &  0.012 & $\langle$B$_5^{(7)}$,M$_2^{(1)}$,M$_2^{(3)}$$\rangle$ &  0.002\\
Al72 & $24d$ &  1.00 & 0.4974(4) & 0.3108(3) & 0.0004(5) &  0.012 & $\langle$B$_5^{(7)}$,M$_2^{(3)}$,M$_2^{(4)}$$\rangle$ &  0.003\\
Al73 & $24d$ &  1.00 & 0.4439(4) & 0.4055(4) & 0.1546(4) &  0.012 & $\langle$B$_5^{(7)}$,M$_2^{(3)}$,M$_2^{(7)}$$\rangle$ &  0.003\\
Al74 & $24d$ &  1.00 & 0.4051(5) & 0.3453(4) & 0.0567(4) &  0.012 & $\langle$B$_5^{(7)}$,M$_2^{(3)}$,M$_2^{(6)}$$\rangle$ &  0.002\\
Al75 & $24d$ &  1.00 & 0.4653(4) & 0.4400(4) & 0.0964(4) &  0.012 & $\langle$B$_5^{(7)}$,M$_2^{(1)}$,M$_2^{(3)}$$\rangle$ &  0.002\\
Cr/Al76 & $4a$ &  0.76/0.24 & 0 & 0 & 0 &  0.015 & $\langle$M$_0^{(1)}$$\rangle$ &  0.000\\
Cr/Al77 & $24d$ &  0.88/0.12 & 0.47737(18) & 0.44032(18) & -0.0002(2) &  0.007 & $\langle$B$_3^{(7)}$,M$_3^{(1)}$$\rangle$ &  0.001\\
Cr/Al78 & $8c$ &  0.96/0.04 & 0.03635(18) & 0.03635(18) & 0.03635(18) &  0.006 & $\langle$B$_3^{(1)}$,M$_3^{(1)}$$\rangle$ &  0.000\\
Cr/Al79 & $4b$ &  0.54/0.46 & 0.50000 & 0 & 0 &  0.007 & $\langle$M$_0^{(2)}$$\rangle$ &  0.000\\
Fe/Pd80 & $8c$ &  0.75/0.25 & 0.46330(13) & 0.46330(13) & 0.46330(13) &  0.002 & $\langle$B$_3^{(2)}$,M$_3^{(2)}$$\rangle$ &  0.000\\
Fe/Pd81 & $24d$ &  0.76/0.24 & 0.46332(15) & 0.03661(15) & 0.03672(15) &  0.013 & $\langle$B$_3^{(5)}$,M$_3^{(2)}$$\rangle$ &  0.000\\
Fe/Pd82 & $24d$ &  0.89/0.11 & 0.34520(19) & 0.44053(19) & 0.0954(2) &  0.017 & $\langle$M$_0^{(3)}$$\rangle$ &  0.001\\
Al83 & $24d$ &  1.00 & 0.3114(5) & 0.4078(6) & 0.0598(5) &  0.048 & $\langle$B$_3^{(6)}$,M$_3^{(3)}$$\rangle$ &  0.004\\
Al84 & $24d$ &  1.00 & 0.3817(6) & 0.4086(7) & 0.0601(7) &  0.081 & $\langle$B$_3^{(7)}$,M$_3^{(3)}$$\rangle$ &  0.004\\
Al85 & $24d$ &  1.00 & 0.3458(5) & 0.4624(5) & 0.0387(5) &  0.045 & $\langle$B$_3^{(7)}$,M$_3^{(3)}$$\rangle$ &  0.003\\
Al86 & $24d$ &  0.45 & 0.3078(8) & 0.4079(8) & 0.1284(8) &  0.008 & $\langle$B$_3^{(3)}$,M$_3^{(3)}$$\rangle$ &  0.005\\
Al86$^\prime$ & $24d$ &  0.55 & 0.2911(6) & 0.4358(6) & 0.1186(6) &  0.008 & $\langle$B$_3^{(3)}$,M$_3^{(3)}$$\rangle$ &  0.038\\
Al87 & $24d$ &  1.00 & 0.3447(5) & 0.4619(5) & 0.1520(5) &  0.025 & $\langle$B$_3^{(4)}$,M$_3^{(3)}$$\rangle$ &  0.003\\
Al88 & $24d$ &  1.00 & 0.4008(6) & 0.4398(6) & 0.1181(6) &  0.077 & $\langle$B$_3^{(7)}$,M$_3^{(3)}$$\rangle$ &  0.004\\
Al89 & $24d$ &  1.00 & 0.3692(6) & 0.4957(6) & 0.0937(6) &  0.075 & $\langle$B$_3^{(1)}$,M$_3^{(3)}$$\rangle$ &  0.005\\
Fe/Pd90 & $24d$ &  0.91/0.09 & 0.4990(2) & 0.19166(13) & 0.0003(2) &  0.012 & $\langle$M$_0^{(4)}$$\rangle$ &  0.001\\
Al91 & $24d$ &  1.00 & 0.2894(5) & 0.0563(5) & 0.0003(5) &  0.041 & $\langle$B$_3^{(4)}$,M$_3^{(4)}$$\rangle$ &  0.004\\
Al92 & $24d$ &  1.00 & 0.2136(5) & 0.4439(6) & 0.0026(6) &  0.054 & $\langle$B$_3^{(6)}$,M$_3^{(4)}$$\rangle$ &  0.004\\
Al93 & $24d$ &  1.00 & 0.3446(4) & 0.0345(4) & 0.0338(4) &  0.007 & $\langle$B$_3^{(5)}$,M$_3^{(4)}$$\rangle$ &  0.003\\
Al94 & $24d$ &  1.00 & 0.1584(6) & 0.4661(6) & 0.0348(6) &  0.065 & $\langle$B$_3^{(5)}$,M$_3^{(4)}$$\rangle$ &  0.005\\
Al95 & $24d$ &  1.00 & 0.4991(6) & 0.4817(5) & 0.2521(5) &  0.048 & $\langle$B$_3^{(7)}$,M$_3^{(4)}$$\rangle$ &  0.005\\
Al96 & $24d$ &  1.00 & 0.4670(8) & 0.4680(8) & 0.3470(8) &  0.105 & $\langle$B$_3^{(2)}$,M$_3^{(4)}$$\rangle$ &  0.006\\
Al97 & $24d$ &  1.00 & 0.469(1) & 0.153(1) & 0.032(1) &  0.152 & $\langle$B$_3^{(5)}$,M$_3^{(4)}$$\rangle$ &  0.007\\
Cr/Al98 & $24d$ &  0.80/0.20 & 0.1531(2) & 0.2497(2) & 0.0947(2) &  0.001 & $\langle$M$_0^{(5)}$$\rangle$ &  0.002\\
Cr/Al99 & $24d$ &  0.72/0.28 & 0.1918(2) & 0.2162(2) & 0.0607(2) &  0.015 & $\langle$B$_3^{(4)}$,M$_3^{(5)}$$\rangle$ &  0.003\\
Al100 & $24d$ &  1.00 & 0.1748(8) & 0.2803(8) & 0.0502(8) &  0.132 & $\langle$B$_3^{(6)}$,M$_3^{(5)}$$\rangle$ &  0.019\\
Cr/Al101 & $24d$ &  0.85/0.15 & 0.13310(19) & 0.1906(2) & 0.0953(2) &  0.007 & $\langle$B$_3^{(1)}$,M$_3^{(5)}$$\rangle$ &  0.001\\
Cr/Al102 & $24d$ &  0.79/0.21 & 0.13219(19) & 0.3076(2) & 0.0952(2) &  0.004 & $\langle$B$_3^{(5)}$,M$_3^{(5)}$$\rangle$ &  0.001\\
Cr/Al103 & $24d$ &  0.77/0.23 & 0.2480(2) & 0.1194(2) & 0.0962(2) &  0.008 & $\langle$B$_3^{(4)}$,M$_3^{(5)}$$\rangle$ &  0.003\\
Al/Cr104 & $24d$ &  0.38/0.27 & 0.191(1) & 0.290(1) & 0.116(1) &  0.160 & $\langle$B$_3^{(3)}$,M$_3^{(5)}$$\rangle$ &  0.016\\
Fe/Pd105 & $24d$ &  0.98/0.02 & 0.1555(2) & 0.75056(19) & 0.0956(3) &  0.021 & $\langle$M$_0^{(6)}$$\rangle$ &  0.001\\
Al106 & $24d$ &  1.00 & 0.3088(4) & 0.2842(4) & 0.1317(4) &  0.014 & $\langle$B$_3^{(3)}$,M$_3^{(6)}$$\rangle$ &  0.002\\
Al/Cr107 & $24d$ &  0.77/0.23 & 0.3087(4) & 0.2844(4) & 0.0586(4) &  0.025 & $\langle$B$_3^{(6)}$,M$_3^{(6)}$$\rangle$ &  0.002\\
Al/Cr108 & $24d$ &  0.83/0.17 & 0.3837(4) & 0.2834(4) & 0.0608(4) &  0.025 & $\langle$B$_3^{(7)}$,M$_3^{(6)}$$\rangle$ &  0.004\\
Al/Cr109 & $24d$ &  0.89/0.11 & 0.4034(4) & 0.2477(4) & 0.1180(4) &  0.023 & $\langle$B$_3^{(6)}$,M$_3^{(6)}$$\rangle$ &  0.003\\
Al110 & $24d$ &  1.00 & 0.3087(5) & 0.2128(5) & 0.0597(5) &  0.034 & $\langle$B$_3^{(4)}$,M$_3^{(6)}$$\rangle$ &  0.001\\
Al111 & $24d$ &  1.00 & 0.3446(5) & 0.2258(5) & 0.1524(4) &  0.034 & $\langle$B$_3^{(3)}$,M$_3^{(6)}$$\rangle$ &  0.003\\
Al112 & $24d$ &  1.00 & 0.3672(6) & 0.1931(6) & 0.0976(6) &  0.067 & $\langle$B$_3^{(5)}$,M$_3^{(6)}$$\rangle$ &  0.003\\
Fe113 & $24d$ &  1.00 & 0.4047(3) & 0.3455(2) & 0.2498(2) &  0.024 & $\langle$M$_0^{(7)}$$\rangle$ &  0.000\\
Al114 & $24d$ &  1.00 & 0.4421(4) & 0.3802(4) & 0.2141(4) &  0.018 & $\langle$B$_3^{(7)}$,M$_3^{(7)}$$\rangle$ &  0.002\\
Al115 & $24d$ &  1.00 & 0.4054(4) & 0.3663(4) & 0.3083(4) &  0.013 & $\langle$B$_3^{(2)}$,M$_3^{(7)}$$\rangle$ &  0.002\\
Al116 & $24d$ &  1.00 & 0.3846(4) & 0.4039(4) & 0.2495(4) &  0.019 & $\langle$B$_3^{(4)}$,M$_3^{(7)}$$\rangle$ &  0.003\\
Al117 & $24d$ &  1.00 & 0.3464(4) & 0.3445(4) & 0.2263(4) &  0.016 & $\langle$B$_3^{(3)}$,M$_3^{(7)}$$\rangle$ &  0.002\\
Al118 & $24d$ &  1.00 & 0.4392(4) & 0.3104(4) & 0.2143(4) &  0.012 & $\langle$B$_3^{(6)}$,M$_3^{(7)}$$\rangle$ &  0.002\\
Al119 & $24d$ &  1.00 & 0.4388(5) & 0.3102(5) & 0.2841(5) &  0.038 & $\langle$B$_3^{(4)}$,M$_3^{(7)}$$\rangle$ &  0.003\\
Al120 & $24d$ &  1.00 & 0.3823(4) & 0.2882(4) & 0.2501(4) &  0.023 & $\langle$B$_3^{(3)}$,M$_3^{(7)}$$\rangle$ &  0.002\\
Pd121 & $24d$ &  1.00 & 0.0379(1) & 0.1538(1) & 0.0334(1) &  0.011 & $\langle$B$_3^{(1)}$,B$_3^{(7)}$,M$_5^{(3)}$$\rangle$ &  0.003\\
Pd122 & $24d$ &  1.00 & 0.15498(9) & 0.34383(9) & 0.0383(1) &  0.001 & $\langle$B$_3^{(5)}$,B$_3^{(6)}$,M$_5^{(5)}$$\rangle$ &  0.003\\
Pd123 & $24d$ &  1.00 & 0.4998(1) & 0.40696(9) & 0.36725(9) &  0.007 & $\langle$B$_3^{(2)}$,B$_3^{(5)}$,M$_5^{(4)}$$\rangle$ &  0.003\\
Pd124 & $24d$ &  1.00 & 0.46348(11) & 0.15417(11) & 0.1535(1) &  0.012 & $\langle$B$_3^{(5)}$,B$_3^{(6)}$$\rangle$ &  0.001\\
Pd125 & $24d$ &  1.00 & 0.4431(1) & 0.49909(11) & 0.2163(1) &  0.013 & $\langle$B$_3^{(4)}$,B$_3^{(7)}$,M$_5^{(4)}$$\rangle$ &  0.004\\
Pd126 & $24d$ &  1.00 & 0.43854(9) & 0.38408(9) & 0.0950(1) &  0.009 & $\langle$B$_3^{(7)}$,B$_3^{(7)}$,M$_5^{(3)}$$\rangle$ &  0.003\\
Pd127 & $24d$ &  1.00 & 0.4613(1) & 0.3446(1) & 0.3433(1) &  0.009 & $\langle$B$_3^{(2)}$,B$_3^{(4)}$,M$_5^{(7)}$$\rangle$ &  0.003\\
Pd128 & $24d$ &  1.00 & 0.25334(9) & 0.38382(9) & 0.0950(1) &  0.006 & $\langle$B$_3^{(3)}$,B$_3^{(6)}$,M$_5^{(3)}$$\rangle$ &  0.004\\
Pd129 & $24d$ &  1.00 & 0.34485(11) & 0.15618(11) & 0.03799(11) &  0.016 & $\langle$B$_3^{(4)}$,B$_3^{(5)}$,M$_5^{(6)}$$\rangle$ &  0.002\\
Pd130 & $24d$ &  1.00 & 0.21331(9) & 0.44028(9) & 0.19174(9) &  0.007 & $\langle$B$_3^{(3)}$,B$_3^{(6)}$$\rangle$ &  0.001\\
Pd131 & $24d$ &  1.00 & 0.09127(9) & 0.36632(9) & 0.00029(11) &  0.011 & $\langle$B$_3^{(5)}$,B$_3^{(5)}$,M$_5^{(4)}$$\rangle$ &  0.005\\
Pd132 & $24d$ &  1.00 & 0.28893(9) & 0.4407(1) & 0.18784(9) &  0.009 & $\langle$B$_3^{(3)}$,B$_3^{(4)}$,M$_5^{(3)}$$\rangle$ &  0.004\\
Pd133 & $24d$ &  1.00 & 0.1904(1) & 0.40743(11) & 0.0554(1) &  0.012 & $\langle$B$_3^{(5)}$,B$_3^{(6)}$,M$_5^{(4)}$$\rangle$ &  0.005\\
Pd134 & $24d$ &  1.00 & 0.19072(9) & 0.40400(9) & 0.13116(9) &  0.005 & $\langle$B$_3^{(3)}$,B$_3^{(5)}$$\rangle$ &  0.001\\
Pd135 & $24d$ &  1.00 & 0.31200(11) & 0.28740(11) & 0.2495(1) &  0.016 & $\langle$B$_3^{(3)}$,B$_3^{(3)}$,M$_5^{(7)}$$\rangle$ &  0.003\\
Pd136 & $24d$ &  1.00 & 0.3089(1) & 0.09278(11) & 0.0566(1) &  0.010 & $\langle$B$_3^{(4)}$,B$_3^{(5)}$,M$_5^{(4)}$,M$_5^{(5)}$$\rangle$ &  0.004\\
Pd137 & $24d$ &  1.00 & 0.1526(1) & 0.1546(1) & 0.03373(11) &  0.013 & $\langle$B$_3^{(1)}$,B$_3^{(4)}$,M$_5^{(3)}$,M$_5^{(5)}$$\rangle$ &  0.003\\
Pd138 & $24d$ &  1.00 & 0.2123(1) & 0.25038(9) & 0.00257(11) &  0.009 & $\langle$B$_3^{(4)}$,B$_3^{(6)}$,M$_5^{(5)}$$\rangle$ &  0.003\\
Pd139 & $24d$ &  1.00 & 0.4390(1) & 0.3074(1) & 0.09494(11) &  0.010 & $\langle$B$_3^{(6)}$,B$_3^{(7)}$,M$_5^{(6)}$$\rangle$ &  0.003\\
Pd140 & $24d$ &  1.00 & 0.1913(1) & 0.0961(1) & 0.0599(1) &  0.009 & $\langle$B$_3^{(1)}$,B$_3^{(4)}$,M$_5^{(5)}$$\rangle$ &  0.001\\
Pd141 & $24d$ &  1.00 & 0.46242(11) & 0.3452(1) & 0.1567(1) &  0.011 & $\langle$B$_3^{(6)}$,B$_3^{(7)}$,M$_5^{(7)}$$\rangle$ &  0.002\\
Pd/Fe142 & $24d$ &  0.74/0.26 & 0.28543(11) & 0.24948(11) & 0.19141(12) &  0.010 & $\langle$B$_3^{(3)}$,B$_3^{(3)}$,M$_5^{(6)}$$\rangle$ &  0.001\\
Fe/Pd143 & $24d$ &  0.51/0.49 & 0.50141(13) & 0.28711(12) & 0.05893(12) &  0.008 & $\langle$B$_3^{(6)}$,B$_3^{(7)}$,M$_5^{(3)}$,M$_5^{(4)}$$\rangle$ &  0.002\\
Pd/Fe144 & $24d$ &  0.63/0.37 & 0.44322(12) & 0.49968(13) & 0.40016(12) &  0.013 & $\langle$B$_3^{(2)}$,B$_3^{(5)}$,M$_5^{(2)}$,M$_5^{(4)}$$\rangle$ &  0.005\\
Pd/Fe145 & $24d$ &  0.78/0.22 & 0.49849(14) & 0.28713(13) & 0.24931(11) &  0.019 & $\langle$B$_3^{(4)}$,B$_3^{(6)}$,M$_5^{(6)}$,M$_5^{(7)}$$\rangle$ &  0.002\\
Pd/Fe146 & $24d$ &  0.84/0.16 & 0.34478(11) & 0.34920(11) & 0.03828(11) &  0.008 & $\langle$B$_3^{(6)}$,B$_3^{(7)}$,M$_5^{(3)}$,M$_5^{(6)}$$\rangle$ &  0.004\\
Pd/Fe147 & $24d$ &  0.68/0.32 & 0.49981(12) & 0.09886(11) & 0.05776(11) &  0.007 & $\langle$B$_3^{(5)}$,B$_3^{(5)}$,M$_5^{(2)}$,M$_5^{(4)}$$\rangle$ &  0.004\\
Pd/Fe148 & $24d$ &  0.55/0.45 & 0.15417(14) & 0.34369(13) & 0.15351(14) &  0.009 & $\langle$B$_3^{(3)}$,B$_3^{(5)}$,M$_5^{(5)}$,M$_5^{(6)}$$\rangle$ &  0.002\\
Pd/Fe149 & $24d$ &  0.55/0.45 & 0.40154(13) & 0.44069(12) & 0.00331(13) &  0.009 & $\langle$B$_3^{(7)}$,B$_3^{(7)}$,M$_5^{(1)}$,M$_5^{(3)}$$\rangle$ &  0.004\\
Pd/Fe150 & $24d$ &  0.57/0.43 & 0.44148(12) & 0.19141(12) & 0.09361(13) &  0.005 & $\langle$B$_3^{(5)}$,B$_3^{(6)}$,M$_5^{(4)}$,M$_5^{(6)}$$\rangle$ &  0.002\\
Pd/Fe151 & $24d$ &  0.64/0.36 & 0.40261(13) & 0.44054(13) & 0.18868(12) &  0.014 & $\langle$B$_3^{(4)}$,B$_3^{(7)}$,M$_5^{(3)}$,M$_5^{(7)}$$\rangle$ &  0.003\\
Pd/Fe152 & $24d$ &  0.53/0.47 & 0.43821(12) & 0.49800(13) & 0.09582(13) &  0.010 & $\langle$B$_3^{(1)}$,B$_3^{(7)}$,M$_5^{(1)}$,M$_5^{(3)}$$\rangle$ &  0.003\\
Pd/Fe153 & $24d$ &  0.55/0.45 & 0.40714(14) & 0.44205(13) & 0.30777(13) &  0.012 & $\langle$B$_3^{(2)}$,B$_3^{(4)}$,M$_5^{(4)}$,M$_5^{(7)}$$\rangle$ &  0.003\\
Fe/Pd154 & $8c$ &  0.72/0.28 & 0.15344(17) & 0.15344(17) & 0.15344(17) &  0.011 & $\langle$B$_3^{(1)}$,M$_5^{(5)}$,M$_5^{(5)}$,M$_5^{(5)}$$\rangle$ &  0.002\\
Fe/Pd155 & $24d$ &  0.58/0.42 & 0.24809(13) & 0.30790(15) & 0.09506(16) &  0.009 & $\langle$B$_3^{(3)}$,B$_3^{(6)}$,M$_5^{(5)}$,M$_5^{(6)}$$\rangle$ &  0.002\\
Pd/Fe156 & $24d$ &  0.64/0.36 & 0.30852(14) & 0.40455(14) & 0.25214(13) &  0.015 & $\langle$B$_3^{(3)}$,B$_3^{(4)}$,M$_5^{(7)}$,M$_5^{(7)}$$\rangle$ &  0.002\\
Pd157 & $24d$ &  0.66 & 0.21000(14) & 0.24990(14) & 0.18636(15) &  0.006 & $\langle$B$_3^{(3)}$,M$_5^{(5)}$$\rangle$ &  0.006\\
Pd157$^\prime$ & $24d$ &  0.34 & 0.1911(3) & 0.2491(3) & 0.1549(3) &  0.009 & $\langle$B$_3^{(3)}$,M$_5^{(5)}$$\rangle$ &  0.042\\
Pd/Fe158 & $24d$ &  0.55/0.45 & 0.40290(15) & 0.25131(13) & 0.19141(15) &  0.015 & $\langle$B$_3^{(3)}$,B$_3^{(6)}$,M$_5^{(6)}$,M$_5^{(7)}$$\rangle$ &  0.002\\
Cr/Al159 & $24d$ &  0.99/0.01 & 0.0963(2) & 0.24942(19) & 0.0002(2) &  0.008 & $\langle$M$_5^{(3)}$,M$_5^{(4)}$,M$_5^{(5)}$$\rangle$ &  0.001\\
Cr/Fe160 & $24d$ &  0.60/0.40 & 0.4035(3) & 0.2486(2) & 0.0013(2) &  0.019 & $\langle$B$_3^{(7)}$,M$_5^{(4)}$,M$_5^{(6)}$,M$_5^{(7)}$$\rangle$ &  0.002\\
Fe/Pd161 & $8c$ &  0.91/0.09 & 0.3445(3) & 0.3445(3) & 0.3445(3) &  0.057 & $\langle$B$_3^{(2)}$,M$_5^{(7)}$,M$_5^{(7)}$,M$_5^{(7)}$$\rangle$ &  0.002\\
Fe/Cr162 & $24d$ &  0.80/0.20 & 0.3470(2) & 0.3444(2) & 0.1534(2) &  0.021 & $\langle$B$_3^{(3)}$,M$_5^{(3)}$,M$_5^{(6)}$,M$_5^{(7)}$$\rangle$ &  0.002\\
Fe/Cr163 & $24d$ &  0.70/0.30 & 0.24888(18) & 0.1904(2) & 0.0955(2) &  0.002 & $\langle$B$_3^{(4)}$,M$_5^{(5)}$,M$_5^{(5)}$,M$_5^{(6)}$$\rangle$ &  0.001\\
Al164 & $24d$ &  1.00 & 0.0602(4) & 0.2148(4) & 0.0364(4) &  0.009 & $\langle$M$_2^{(3)}$,M$_2^{(4)}$,M$_2^{(5)}$$\rangle$ &  0.002\\
Al165 & $24d$ &  1.00 & 0.4990(4) & 0.3805(4) & 0.3086(4) &  0.016 & $\langle$M$_2^{(4)}$,M$_2^{(6)}$,M$_2^{(7)}$$\rangle$ &  0.002\\
Al166 & $24d$ &  1.00 & 0.2270(3) & 0.4421(4) & 0.0947(4) &  0.010 & $\langle$M$_2^{(3)}$,M$_2^{(4)}$$\rangle$ &  0.001\\
Al167 & $24d$ &  1.00 & 0.1915(4) & 0.5005(4) & 0.1184(4) &  0.013 & $\langle$M$_2^{(4)}$,M$_2^{(5)}$$\rangle$ &  0.001\\
Al168 & $24d$ &  1.00 & 0.2488(4) & 0.4763(4) & 0.1540(4) &  0.008 & $\langle$M$_2^{(3)}$,M$_2^{(5)}$$\rangle$ &  0.002\\
Al169 & $24d$ &  1.00 & 0.4037(4) & 0.3463(4) & 0.1326(4) &  0.014 & $\langle$M$_2^{(3)}$,M$_2^{(6)}$,M$_2^{(7)}$$\rangle$ &  0.001\\
Al170 & $8c$ &  1.00 & 0.1900(4) & 0.1900(4) & 0.1900(4) &  0.000 & $\langle$M$_2^{(5)}$,M$_2^{(5)}$,M$_2^{(5)}$$\rangle$ &  0.002\\
Al171 & $24d$ &  1.00 & 0.2497(4) & 0.2140(5) & 0.1531(5) &  0.022 & $\langle$M$_2^{(5)}$,M$_2^{(5)}$,M$_2^{(6)}$$\rangle$ &  0.002\\
Al172 & $8c$ &  1.00 & 0.3092(5) & 0.3092(5) & 0.3092(5) &  0.029 & $\langle$M$_2^{(7)}$,M$_2^{(7)}$,M$_2^{(7)}$$\rangle$ &  0.000\\
Al173 & $24d$ &  0.53 & 0.2086(8) & 0.1610(8) & 0.1295(8) &  0.029 & $\langle$M$_2^{(5)}$$\rangle$ &  0.009\\
Al173$^\prime$ & $24d$ &  0.47 & 0.190(1) & 0.1909(9) & 0.1204(9) &  0.029 & $\langle$M$_2^{(5)}$$\rangle$ &  0.003\\
Al174 & $24d$ &  0.72 & 0.0964(5) & 0.2698(5) & 0.0594(5) &  0.011 & $\langle$M$_2^{(4)}$$\rangle$ &  0.003\\
Al174$^\prime$ & $24d$ &  0.54 & 0.0515(6) & 0.2840(6) & 0.0041(6) &  0.011 & $\langle$M$_2^{(5)}$$\rangle$ &  0.033\\
Al175 & $24d$ &  0.61 & 0.2834(6) & 0.5380(6) & 0.1334(6) &  0.017 & $\langle$M$_2^{(3)}$$\rangle$ &  0.004\\
Al175$^\prime$ & $24d$ &  0.39 & 0.3139(9) & 0.4941(9) & 0.1078(9) &  0.017 & $\langle$M$_2^{(5)}$$\rangle$ &  0.013\\
Al176 & $24d$ &  0.69 & 0.2869(7) & 0.1605(7) & 0.1314(7) &  0.038 & $\langle$M$_2^{(6)}$$\rangle$ &  0.006\\
Al176$^\prime$ & $24d$ &  0.31 & 0.1151(15) & 0.3102(16) & 0.1911(16) &  0.038 & $\langle$M$_2^{(5)}$$\rangle$ &  0.003\\
Al177 & $24d$ &  0.45 & 0.2306(11) & 0.250(1) & 0.0973(12) &  0.036 & $\langle$M$_2^{(6)}$$\rangle$ &  0.004\\
Al177$^\prime$ & $24d$ &  0.55 & 0.2659(9) & 0.2487(8) & 0.0952(9) &  0.036 & $\langle$M$_2^{(5)}$$\rangle$ &  0.007\\
Al178 & $24d$ &  0.52 & 0.3678(11) & 0.2852(11) & 0.1503(11) &  0.057 & $\langle$M$_2^{(7)}$$\rangle$ &  0.004\\
Al178$^\prime$ & $24d$ &  0.48 & 0.3790(11) & 0.3079(12) & 0.1827(11) &  0.057 & $\langle$M$_2^{(6)}$$\rangle$ &  0.009\\
Al179 & $24d$ &  0.64 & 0.0964(7) & 0.3458(7) & 0.3218(6) &  0.029 & $\langle$M$_2^{(3)}$$\rangle$ &  0.001\\
Al179$^\prime$ & $24d$ &  0.36 & 0.0996(12) & 0.3508(12) & 0.3774(11) &  0.029 & $\langle$M$_2^{(6)}$$\rangle$ &  0.012\\
Al180 & $24d$ &  0.62 & 0.5259(8) & 0.3437(8) & 0.2504(7) &  0.031 & $\langle$M$_2^{(7)}$$\rangle$ &  0.004\\
Al180$^\prime$ & $24d$ &  0.38 & 0.3484(12) & 0.2503(11) & -0.0147(12) &  0.031 & $\langle$M$_2^{(6)}$$\rangle$ &  0.008\\
Al181 & $24d$ &  0.48 & 0.3824(12) & 0.3813(12) & 0.1876(12) &  0.059 & $\langle$M$_2^{(3)}$$\rangle$ &  0.003\\
Al181$^\prime$ & $24d$ &  0.52 & 0.3683(11) & 0.4001(11) & 0.1561(11) &  0.059 & $\langle$M$_2^{(7)}$$\rangle$ &  0.005\\
Al182 & $24d$ &  0.57 & 0.347(1) & 0.367(1) & 0.286(1) &  0.058 & $\langle$M$_2^{(7)}$$\rangle$ &  0.002\\
Al182$^\prime$ & $24d$ &  0.43 & 0.3129(13) & 0.3786(13) & 0.3085(13) &  0.058 & $\langle$M$_2^{(7)}$$\rangle$ &  0.005\\
Al183 & $24d$ &  0.23 & 0.5371(17) & 0.2480(16) & 0.0218(17) &  0.015 & $\langle$M$_2^{(3)}$$\rangle$ &  0.002\\
Al183$^\prime$ & $24d$ &  0.51 & 0.5538(7) & 0.2776(7) & 0.0321(7) &  0.015 & $\langle$M$_2^{(4)}$$\rangle$ &  0.011\\
Al184 & $24d$ &  0.76 & 0.4253(4) & 0.0003(5) & 0.0002(5) &  0.009 & $\langle$M$_2^{(4)}$$\rangle$ &  0.002\\
Al184$^\prime$ & $24d$ &  0.24 & -0.0014(14) & 0.4007(12) & -0.0007(15) &  0.009 & $\langle$M$_2^{(2)}$$\rangle$ &  0.019\\
Al185 & $24d$ &  0.66 & 0.4365(9) & 0.2128(9) & 0.0366(9) &  0.067 & $\langle$M$_2^{(6)}$$\rangle$ &  0.005\\
Al185$^\prime$ & $24d$ &  0.34 & 0.3913(17) & 0.2214(17) & 0.0618(17) &  0.067 & $\langle$M$_2^{(4)}$$\rangle$ &  0.015\\
Al186 & $24d$ &  0.61 & 0.4403(6) & 0.4773(5) & 0.0363(6) &  0.007 & $\langle$M$_2^{(3)}$$\rangle$ &  0.001\\
Al186$^\prime$ & $24d$ &  0.39 & 0.4193(9) & 0.4689(8) & 0.0510(9) &  0.007 & $\langle$M$_2^{(1)}$$\rangle$ &  0.018\\
Al187 & $24d$ &  0.59 & 0.4658(9) & 0.4381(9) & 0.2866(9) &  0.050 & $\langle$M$_2^{(7)}$$\rangle$ &  0.004\\
Al187$^\prime$ & $24d$ &  0.41 & 0.4423(13) & 0.4010(13) & 0.2735(12) &  0.050 & $\langle$M$_2^{(4)}$$\rangle$ &  0.004
\end{longtable*}



\begin{acknowledgments}
The authors are grateful to Satoshi Ohashi for technical assistance during the TEM and SEM observations. They are also indebted to Hisanori Yamane for his guidance and helpful comments at various stages of the single crystal X-ray diffraction measurement and analysis. The single crystal X-ray diffraction measurement was performed by Akira Sato, whose courtesy and effort in making the high quality data available is highly appreciated. A large part of the figures (Figures 5 to 15) have been drawn using a three-dimensional visualization software package for crystal structures, VESTA ver.2.9 \cite{VESTA}.
\end{acknowledgments}

\bibliography{alpdtm.bib}






\end{document}